\documentclass[aps,prd,reprint,superscriptaddress,flushbottom,longbibliography]{revtex4-1}

\usepackage{general_latex}

\begin{document}

\title{Quantum sine-Gordon dynamics on analogue curved spacetime in a weakly imperfect scalar Bose gas}
\author{T.\,J.~Volkoff}
\email{volkoff@snu.ac.kr}
\author{Uwe R. Fischer}
\email{uwe@phya.snu.ac.kr}
\affiliation{Seoul National University,   Department of Physics and Astronomy \\  Center for Theoretical Physics, 
Seoul 08826, Korea}

\begin{abstract}
Using the coherent state functional integral expression of the partition function, we show that the sine-Gordon model on an analogue curved spacetime arises as the effective quantum field theory for phase fluctuations of a weakly imperfect Bose gas on an incompressible background superfluid flow when these fluctuations are restricted to a subspace of the single-particle Hilbert space. We consider bipartitions of the single-particle Hilbert space relevant to experiments on ultracold bosonic atomic or molecular gases, including, e.g., restriction to high- or low-energy sectors of the dynamics and spatial bipartition corresponding to tunnel-coupled planar Bose gases. By assuming full unitary quantum control in the low-energy subspace of a trapped gas, we show that (1) appropriately tuning the particle number statistics of the lowest-energy mode partially decouples the low- and high-energy sectors, allowing any low-energy single-particle wave function to define a background for sine-Gordon dynamics on curved spacetime and (2) macroscopic occupation of a quantum superposition of two states of the lowest two modes produces an analogue curved spacetime depending on two background flows, with respective weights continuously dependent on the corresponding weights of the superposed quantum states.
\end{abstract}
\maketitle

\section{Introduction}

The weakly imperfect Bose gas (WIBG) represents a paradigmatic quantum system supporting excitations that propagate in an analogue curved spacetime (ACS) \cite{barcelorev}. Recent progress in quantum control and measurement of optically trapped ultracold alkali gases suggests that several aspects of quantum field dynamics on analogue curved spacetimes are accessible to experimental studies  in dilute ultracold Bose gases. With the advances in experimental precision, such effects as, e.g., Hawking radiation \cite{steinhauer1} in a black hole laser \cite{BHlaser}, Sakharov oscillations \cite{Chin,Sakharov}, as well as the analogue of cosmological particle production (dynamical Casimir effect) \cite{Jaskula,CPP} have been detected. 

To observe the interplay between control of the quantum state of the WIBG and the quantum dynamics on ACS of a relevant effective field, an ideal experiment should be able to address both the mode occupation statistics of the gas and the dynamics of the effective field. An example protocol utilizing the WIBG as a quantum simulator of quantum field theory on curved spacetime could entail: (1) preparation of sufficiently long-lived nonclassical states of a subset of single-particle modes of the WIBG, (2) manipulation of the effective quantum field propagating in ACS, e.g., quenching the effective field, and (3) inference of properties of the effective field through measurements of the Bose gas. However, when a subset of WIBG modes has been prepared in a given quantum state, it is not clear what the effective ACS dynamics of quantum fluctuations of the remaining modes will be. The dynamics depends on the coupling between the mode sectors and the effective potential arising in the fluctuating sector. In particular, the resulting dynamics may not be that of a free particle on curved spacetime, i.e., may not give rise simply to the wave equation $\square_{g}\theta=0$, where $\square_{g}$ is the Laplace-Beltrami operator.

Among continuum quantum models exhibiting a nonlinear interaction in the field operators, the quantum sine-Gordon model is notable for its exact solubility and mapping to a fermionic model in one time dimension and one space dimension (i.e., (1+1)-D) \cite{izergin,colemanthirr} and its wide applicability in condensed matter systems exhibiting global U(1) symmetry, e.g., long \cite{paterno} and annular \cite{ustinov} Josephson junctions in superconducting circuits. In the context of bosons interacting via $s$-wave scattering, it is known that two tunnel-coupled (1+1)-D  WIBG systems in one space and one time dimension exhibit sine-Gordon dynamics of the relative phase between the systems in the limit of Luttinger liquid dynamics \cite{demlersine}. Furthermore, the (1+1)-D sine-Gordon model in an expanding spacetime described by a Friedmann-Robertson-Walker metric has been studied by including a time-dependent mass term arising from time-dependent tunneling between two WIBGs in the Luttinger hydrodynamic limit \cite{marquardtcosmo}.  In the case of (2+1)-D and (3+1)-D, a candidate system for simulating quantum sine-Gordon dynamics on ACS is, however, lacking. Below, we provide examples of such systems which can, in principle, be experimentally realized with ultracold bosonic quantum gases.

In this paper, we show that a general procedure consisting of (1) partitioning the single-particle modes of the WIBG into two subsets $J_{L}$ and $J_{H}$ and (2) pinning the dynamics of one subset, e.g., $J_{L}$, to its action-extremizing equation of motion (solutions of which self-consistently define the single-particle states of the $J_{L}$ sector and, therefore, the modes comprising the vacuum for the $J_{H}$ quantum fluctuations), allows the phase fluctuations of the $J_{H}$ field to be described as bosons propagating on a curved spacetime in a sine-Gordon potential. Sections \ref{sec:jhoneloop} and \ref{sec:jhhighloop} contain the general derivation. We discuss the sine-Gordon equation on ACS in Sec. \ref{sec:sgeqonacs}. Proceeding to example systems, we first consider in Sec. \ref{sec:engineer} the sine-Gordon dynamics on ACS after preparation of the lowest mode in a coherent state (equivalent to the zero mode $c$-number substitution of Bogoliubov) and in a superposition of coherent states of opposite phase. In Sec. \ref{sec:interplane}, we consider a spatial bipartition of the single-particle modes in tunnel-coupled (2+1)-D planes of two WIBGs. Section \ref{sec:superposacs} contains a derivation of the sine-Gordon dynamics on ACS when a WIBG system is projected to a subspace of bosonic Fock space in which the lowest two modes are prepared in a macroscopic superposition state. This extreme case highlights some of the unusual properties of ACS supported on a nonclassical vacuum, departing significantly from the ACS arising from a single semiclassical background field.

\section{sine-Gordon dynamics on ACS\label{sec:sgdyn}}

We consider a single-particle Hilbert space spanned by an orthonormal basis $\lbrace \ket{\varphi_{j}}\rbrace_{j\in J}$ and a nonrelativistic quantum field $\hat{\psi}(x)=\sum_{j\in J}\varphi_{j}(x)\hat{a}_{j}$, where $J$ is an index set, $\varphi_{j}(x) \in L^{2}(\Omega \subset \mathbb{R}^{3})$, and $\hat{a}_{j}$ is the bosonic annihilation operator. The finite volume of the trap containing the WIBG is labeled $\vert \Omega \vert$. The normal ordered weakly imperfect Bose gas Hamiltonian in the presence of a U(1) gauge field $v(x)$ describing, e.g., a rotation, Galilei boost, or other background velocity field, is given by (suppressing the spatial dependence of the field operators):
\begin{eqnarray}
\hat{H} &=& \int_{\Omega}d^{3}x \, {\hbar^{2}\over 2m}\overline{D} \hat{\psi}^{\dagger}D\hat{\psi} 
+(mV_{\text{ext}}(x) - \mu)\hat{\psi}^{\dagger}\hat{\psi} \nonumber \\ &+&{V_{0}\over 2}\hat{\psi}^{\dagger \, 2}\hat{\psi}^{2}\nn
\label{eqn:wibgham}\end{eqnarray} where $D := \nabla -i{m\over \hbar}v(x)$ is the U(1) covariant derivative, $V_{\text{ext}}(x)$ is an external one-body potential, $V_{0}=(4\pi \hbar^2/m)a_s$ is the contact interaction coupling, with $a_s$ the $s$-wave scattering length,  
and $m$ is taken as the bare mass of the atomic or molecular constituent of the gas. The temperature-dependent chemical potential $\mu$ is defined such that $N = -{1\over \beta}\del_{\mu}\log \tr e^{-\beta \hat{H}}$, with $N$ the average number of gas atoms and $\beta$ the inverse temperature.

In this section, we aim to show that when $J$ is partitioned into two subsets $J_{H}$ and $J_{L}$ and the dynamics of one subset is pinned to a self-consistent equation of motion, the complementary subset exhibits sine-Gordon dynamics on ACS. The sine-Gordon mass will be proportional to $V_{0}n_{L,0}n_{H,0}/ \sqrt{-g}$ where $V_{0}$ is the interaction strength, $n_{L,0}n_{H,0}$ is the product of local number densities in the $J_{L}$ and $J_{H}$ sector, and $\sqrt{-g}:= \sqrt{-\det g_{\mu \nu}}$, where $g_{\mu\nu}$ is the ACS metric in Eq. (\ref{eqn:covar}). To demonstrate these features, we construct the coherent state path integral \cite{negele} for the partition function instead of approximating the operator equations of motion for the weakly imperfect Bose gas. This choice allows to derive the effective action for the phase fluctuations of the quantum field in the $J_{H}$ sector without having to explicitly quantize the phase fluctuation field operator on the ACS. The present approach also allows to more easily consider the effect of nonzero temperature on the contribution of quantum fluctuations to the resulting effective action. 

The derivation of sine-Gordon dynamics on ACS in the $J_{H}$ sector proceeds as follows: we first show that the one-loop effective dynamics of phase fluctuations in the $J_{H}$ sector is that of a massive Klein-Gordon field on ACS (Sec. \ref{sec:modepartitionsubsec} presents the partitioning of the single-particle modes and Sec. \ref{sec:onelooppartitioned} presents the massive Klein-Gordon dynamics on ACS). In Sec. \ref{sec:jhhighloop}, we sum the higher-loop contributions to the dynamics in the $J_{H}$ sector to derive the sine-Gordon dynamics on ACS  and in Sec. \ref{sec:sgeqonacs}, we further analyze the sine-Gordon equation arising from the effective dynamics.

\subsection{ACS in the $J_{H}$ sector at one-loop order\label{sec:jhoneloop}}
\subsubsection{Partition of single-particle modes \label{sec:modepartitionsubsec}}
When the field operator is decomposed as $\hat{\psi} = \hat{\psi}_{L} + \hat{\psi}_{H}$ where $\hat{\psi}_{L(H)}:= \sum_{j\in J_{L(H)}}\varphi_{j}(x)\hat{a}_{j}$ and $J=J_{L}\sqcup J_{H}$ is a bipartition of the set of modes, one can verify that $\hat{\psi}_{L(H)}(x)\ket{\lbrace \phi \rbrace } = \psi_{L(H)}[\phi]\ket{\lbrace \phi \rbrace }$ where $\ket{\lbrace \phi \rbrace } := 
\exp\left[-{1 \over 2}\int_{\Omega}d^{3}x\, \vert \phi (x) \vert^{2}\right]
\exp\left[\int_{\Omega}d^{3}x\, \phi(x)\hat\psi^{\dagger}(x)\right] \ket{\text{VAC}}$ is the normalized field coherent state and where $\psi_{L(H)}[\phi]:= \left(\sum_{j\in J_{L(H)}} \int_{\Omega}d^{3}x'\, \phi(x')\overline{\varphi_{j}(x')}\varphi_{j}(x) \right)$ is the projection of the function $\phi(x)$ onto the space spanned by the single-particle wave functions in $J_{L}$ or $J_{H}$. It follows that $\hat{\psi}\ket{\lbrace \phi \rbrace }= \left( \psi_{L}[\phi] + \psi_{H}[\phi]\right)\ket{\lbrace \phi \rbrace}$. Therefore, the action $S$ appearing in the imaginary time coherent state path integral 
for the partition function $Z(\beta)=\text{tr}\left[e^{-\beta\hat{H}}\right] = \int \prod_{j=L,H} \mathcal{D}[\psi_{j},\overline{\psi_{j}}]e^{-S}$ can be written as follows:
\begin{eqnarray}
S&=&\int_{0}^{\beta \hbar}{d\tau \over \hbar}\int_{\Omega}d^{3}x\, \left[\vphantom{\sum} (\overline{\psi_{L}}+\overline{\psi_{H}})\hbar\del_{\tau}(\psi_{L} + \psi_{H}) \right. \nonumber \\  &+& \left. H[\overline{\psi_{L}}+\overline{\psi_{H}},\psi_{L}+\psi_{H}] \vphantom{\sum}\right]
\label{eqn:actionfull}
\end{eqnarray}
where $H[\overline{\psi_{L}}+\overline{\psi_{H}},\psi_{L}+\psi_{H}]$ is given by the formal substitution $\hat{\psi} \rightarrow \psi_{L}+\psi_{H}$, $\hat{\psi}^{\dagger}\rightarrow \overline{\psi_{L}}+\overline{\psi_{H}}$ in Eq.(\ref{eqn:wibgham}), and we have shortened the symbols for the field eigenvalues to $\psi_{L}$ and $\psi_{H}$, respectively. To show how the bipartitioning of modes leads to  sine-Gordon dynamics on ACS, we now arbitrarily choose $J_{H}$ as the support modes for the phase fluctuations. We require that the field $\psi_{L}$ satisfy the generalized imaginary time Gross-Pitaevskii equation \cite{zarembabook}
\begin{eqnarray}&&\hbar\del_{\tau}\psi_{L} -{\hbar^{2}\over 2m}D^{2}\psi_{L} + (mV_{\text{ext}}(x)-\mu + V_{0}\vert \psi_{H} \vert^{2})\psi_{L} \nonumber \\ &&+ V_{0}\vert \psi_{L} \vert^{2}\psi_{L} = 0
\label{eqn:statphase}
\end{eqnarray}
and that $\overline{\psi_{L}}$ satisfy the associated adjoint field equation

\begin{eqnarray}&&-\hbar\del_{\tau}\overline{\psi_{L}} -{\hbar^{2}\over 2m}\overline{D}^{2}\overline{\psi_{L}}+ (mV_{\text{ext}}(x)-\mu + V_{0}\vert \psi_{H} \vert^{2})\overline{\psi_{L}}  \nonumber \\ &&+ V_{0}\vert \psi_{L} \vert^{2}\overline{\psi_{L}} = 0 .
\label{eqn:statphaseadj}
\end{eqnarray}
These equations are defined on a subspace of $L^{2}(\Omega)$ spanned by the wave functions $\lbrace \varphi_{j}(x) \rbrace_{j\in J_{L}}$. To obtain solutions to the equations above, $\vert \psi_{H}\vert^{2}$ must be calculated at each order in perturbation theory (e.g., at tree order from Eq.(\ref{eqn:euler}) below, giving $\vert \psi_{H}\vert^{2} =n_{H,0}$), substituted into the self-consistent equations Eq.(\ref{eqn:statphase}) and Eq.(\ref{eqn:statphaseadj}), and subsequently solved.  Note that by demanding that the field with support in $J_{L}$ satisfy the generalized Gross-Pitaevskii equation, we are neglecting quantum fluctuations in this sector (i.e., there is no longer a path integral over the field $\psi_{L}$ in $Z(\beta)$, only a sum over solutions $\psi_{L,0}$ to Eq.(\ref{eqn:statphase})). Equivalently, we must restrict to a state of the weakly imperfect Bose gas such that $\hat{\psi} \approx \hat{\psi}_{H} + \langle \hat{\psi}_{L} \rangle = \hat{\psi}_{H} + \psi_{L,0}$ is a valid approximation for the field operator. For the definition of ACS, it will also be important that $\langle \hat{\psi}_{H} \rangle \neq 0$ in this state. Such a state can occur in a nonuniform Bose gas with large occupation number in both the $J_{L}$ sector and $J_{H}$ sector. If the set $J$ is ordered by, e.g., energy values, and $J_{L}$ corresponds to the low-energy modes, this section can be considered as a derivation of the effective field theory of the high-energy phase fluctuations when the low-energy sector is pinned to tree-level. In Secs. \ref{sec:engineer} and \ref{sec:superposacs} we show that the complication arising from requiring a self-consistent solution of the above equation can be removed by preparing the $J_{L}$ modes in an appropriate nonclassical state.

To proceed with deriving the action in the $J_{H}$ sector, we substitute solutions $\psi_{L,0}$ and $\overline{\psi_{L,0}}$ of Eq.(\ref{eqn:statphase}) and Eq.(\ref{eqn:statphaseadj}) for $\psi_{L}$ and $\overline{\psi_{L}}$, respectively, into Eq.(\ref{eqn:actionfull}). Multiplying Eq.(\ref{eqn:statphase}) and Eq.(\ref{eqn:statphaseadj}) by $\overline{\psi_{H}}$ and $\psi_{H}$, respectively, one finds that all monomials in the fields and their derivatives involving both $\psi_{L,0}$ and $\psi_{H}$ in Eq.(\ref{eqn:actionfull}) vanish, except for $\left( \overline{\psi_{H}}^{2}\psi_{L,0}^{2} + c.c.\right)$ and $\vert \psi_{L,0} \vert^{2}\vert \psi_{H}\vert^{2}$. Therefore, the action in Eq.(\ref{eqn:actionfull}) with the $J_{L}$ fields pinned to their stationary phase configurations simplifies to $S_{L,0} + S_{H}$, where
\begin{eqnarray}
S_{H} &=&\int_{0}^{\beta \hbar}{d\tau \over \hbar}\int_{\Omega}d^{3}x \, \left[\vphantom{{A\over B}} \overline{\psi_{H}}\hbar \del_{\tau}\psi_{H} +{\hbar^{2}\over 2m}\overline{D}\,\overline{\psi_{H}}D\psi_{H} \right. \nonumber \\ &+& \left.  \left( mV_{\text{ext}}-\mu \right)\vert \psi_{H}\vert^{2} +{V_{0}\over 2}\left( \overline{\psi_{H}}^{2}\psi_{L,0}^{2} + c.c.\right) \right. \nonumber  \\ &+& \left.   2V_{0}\vert \psi_{L,0} \vert^{2}\vert \psi_{H}\vert^{2} + {V_{0}\over 2}\vert \psi_{H} \vert^{4}\vphantom{{A\over B}}\right]
\end{eqnarray}
and where $S_{L,0}$ is the energy functional \begin{eqnarray}S_{L,0}&=&\int_{0}^{\beta \hbar}{d\tau \over \hbar}\int_{\Omega}d^{3}x \, \left[\vphantom{{A\over B}} \overline{\psi_{L,0}}\hbar \del_{\tau}\psi_{L,0} +{\hbar^{2}\over 2m}\overline{D}\,\overline{\psi_{L,0}}D\psi_{L,0} \right. \nonumber \\ &+& \left.  \left( mV_{\text{ext}}-\mu \right)\vert \psi_{L,0}\vert^{2} +{V_{0}\over 2}\vert \psi_{L,0} \vert^{4} \vphantom{{A\over B}} \right]\label{eqn:lowlandau} \end{eqnarray}
which depends only on the solutions $\psi_{L,0}$, $\overline{\psi_{L,0}}$ to  Eq.(\ref{eqn:statphase}) and Eq.(\ref{eqn:statphaseadj}).

The dynamics of phase fluctuations in the $J_{H}$ sector can be derived by first writing the stationary phase solution of the $J_{L}$ sector in the polar form $\psi_{L,0} = \sqrt{n_{L,0}}e^{i\theta_{L,0} / \hbar}$ and, similarly, performing the change of field variables $\psi_{H}=\sqrt{n_{H}}e^{i\theta_{H} / \hbar}$ in $J_{H}$. The resulting approximate partition function, containing a functional integration over fields $n_{H}$ and $\theta_{H}$ and a sum over all solutions $\psi_{L,0}$ of the generalized Gross-Pitaevskii equation with appropriate boundary conditions imposed on the domain $\Omega \times [0,\beta\hbar]$, is 
\begin{eqnarray}
Z(\beta)&\approx &\sum_{\psi_{L,0}}\int \mathcal{D}[n_{H}]\mathcal{D}[\theta_{H}] e^{-\left( S_{L,0}+S_{H} \right) }
\label{eqn:partsum}
\end{eqnarray} 
where the high-energy part $S_{H}$ of the action is given by 
\begin{eqnarray}
S_{H}&=&\int_{0}^{\beta \hbar}{d\tau \over \hbar}\int_{\Omega} \left[\vphantom{{V_{0}\over 2}} in_{H}\del_{\tau}\theta_{H} + (mV_{\text{ext}} - \mu + 2V_{0}n_{L,0})n_{H} \right. \nonumber  \\ &+& \left. {\hbar^{2} \over 8m}n_{H}^{-1}\nabla n_{H} \cdot \nabla n_{H} + {1 \over 2m}n_{H}\nabla \theta_{H} \cdot \nabla \theta_{H} \right. \nonumber  \\   &-& \left.  n_{H} v\cdot \nabla \theta_{H} + {m\over 2}n_{H}v\cdot v \right. \nonumber  \\   &+& \left. V_{0}n_{H}n_{L,0}\cos((2\theta_{H}-2\theta_{L,0})/\hbar)) + {V_{0}\over 2}n_{H}^{2} \vphantom{\sum}\right]  .
\label{eqn:polarform} 
\end{eqnarray}
Here, we note that the temperature $\beta$ enters not only the solution pair $(n_{L,0}(\tau ,x) , \theta_{L,0}(\tau,x))$,  which is periodic on $\tau \in [0,\beta\hbar]$, but also defines the equilibrium state in the $J_{H}$ sector. Further discussion of the effect of nonzero temperature on the effective theory of phase fluctuations on ACS is provided in the Appendix.

\subsubsection{Massive Klein-Gordon dynamics at one-loop order and phase-matching condition\label{sec:onelooppartitioned}}
We proceed by assuming that the ``quantum potential'' term ${\hbar^{2} \over 8m}n_{H}^{-1}\nabla n_{H} \cdot \nabla n_{H}$ in Eq.(\ref{eqn:polarform}) is negligible \footnote{This condition is one of four assumptions that, taken together, are sufficient for the derivation of a metric tensor that defines the local spacetime on which phase fluctuations of the WIBG propagate. The four assumptions are discussed in Appendix A.}, which is an extensively studied (long wavelength) 
approximation \cite{BLVBEC}. The action $S_{H}$ is now expanded to one-loop order about a solution pair $( n_{H,0},\theta_{H,0} )$, of the stationary phase equations $\delta S_{H} / \delta n_{H} = 0$ and $\delta S_{H} / \delta \theta_{H} = 0$ in the same way as in the general derivation in Appendix A. In imaginary time, the stationary phase equations are 

\begin{eqnarray}
{\delta S_{H} \over \delta n_{H}} = 0 &\Leftrightarrow& i\del_{\tau}\theta_{H} +{1\over 2m}\left( \nabla \theta_{H} - mv\right)\cdot \left(\nabla \theta_{H} - mv \right)  \nonumber \\ &+&   mV_{\text{ext}}(x) - \mu  + V_{0}n_{H}\nonumber \\
&=& -2V_{0}n_{L,0}- V_{0}n_{L,0}\cos ((2\theta_{H}-2\theta_{L,0})/\hbar)) , \nn
{\delta S_{H} \over \delta \theta_{H}} = 0 &\Leftrightarrow& -i\del_{\tau}n_{H} - {1\over m}\nabla \cdot \left( n_{H} \left( \nabla \theta_{H} - mv \right) \right) \nonumber \\ & =& {2V_{0}n_{H}n_{L,0}\over \hbar}\sin((2\theta_{H}-2\theta_{L,0})/\hbar)).
\label{eqn:euler}
\end{eqnarray}
These equations are the internal Josephson equation and the mass continuity equation, respectively, in the $J_{H}$ sector and have solution pairs labeled $\theta_{H,0}$, $n_{H,0}$. It is intriguing to note that while the backreaction on the mean field phase in the $J_{H}$ sector due to the mean field phase of the $J_{L}$ sector never vanishes, the backreaction on the mean field $n_{H,0}$ due to the difference in the mean field phases can vanish for certain solutions of the stationary phase equations. From Eq.(\ref{eqn:polarform}) and Eq.(\ref{eqn:euler}), it is clear that particle number conservation in the $J_{H}$ sector is satisfied at both the highest energy configurations ${2\theta_{H,0} -2\theta_{L,0} \over \hbar} = 2k\pi$ and the lowest energy configurations ${2\theta_{H,0} -2\theta_{L,0} \over \hbar} = (2k+1)\pi$, $k \in \mathbb{Z}$, of the background phases. 

The functional Hessian of $S_{H}$ evaluated at $\theta_{H,0}$, $n_{H,0}$ is given by 
\begin{widetext}
\begin{eqnarray}
\delta^{2} S_{H} \over \delta n_{H}(x,\tau)\delta n_{H}(x',\tau') &=& V_{0}  \delta(x-x')\delta(\tau-\tau ') , \nonumber \\
{\delta^{2} S_{H} \over \delta n_{H}(x,\tau)\delta \theta_{H}(x',\tau')} &=&  -i \del_{\tau}\delta(x-x')\delta(\tau-\tau ') -{1\over m}\nabla \theta_{H,0}\cdot \nabla \delta(x-x')\delta(\tau-\tau ') - {1\over m}\delta(x-x')\delta(\tau-\tau ')\nabla^{2}\theta_{H,0}  \nonumber  \\ &{}&   +  v\cdot \nabla \delta(x-x')\delta(\tau-\tau ') -2{V_{0}\over \hbar}n_{H,0}\sin\left({2\theta_{H,0}-2\theta_{L,0} \over \hbar}\right) \delta(x-x')\delta(\tau-\tau ') , \nonumber \\ {\delta^{2} S_{H} \over \delta \theta_{H}(x,\tau)\delta n_{H}(x',\tau')}&=&i \del_{\tau}\delta(x-x')\delta(\tau-\tau ') + {1\over m} \nabla \theta_{H,0} \cdot \nabla \delta(x-x')\delta(\tau-\tau ') -  v\cdot \nabla \delta(x-x')\delta(\tau-\tau ') \nonumber \\ &{}& -2{V_{0}\over \hbar}n_{H,0}\sin\left({2\theta_{H,0}-2\theta_{L,0} \over \hbar}\right)\delta(x-x')\delta(\tau-\tau '), \nonumber \\ {\delta^{2} S_{H} \over \delta \theta_{H}(x,\tau)\delta \theta_{H}(x',\tau')} &=& -{1\over m}\nabla n_{H,0} \cdot \nabla \delta(x-x')\delta(\tau-\tau ') - {1\over m}n_{H,0}\nabla^{2}\delta(x-x')\delta(\tau-\tau ') \nonumber \\ &{}& -4{V_{0} \over \hbar^{2}}n_{L,0}n_{H,0}\cos\left({2\theta_{H,0}-2\theta_{L,0} \over \hbar}\right)\delta(x-x')\delta(\tau-\tau ').
\label{eqn:highoneloop}
\end{eqnarray}
\end{widetext}

 At this point, it is useful to note that at one-loop order, the action is given schematically by
\begin{eqnarray}
S&=&S_{L,0} + S_{H}[n_{H,0},\theta_{H,0};n_{L,0},\theta_{L,0}] \nonumber \\ &+&{1\over 2!}\int (n_{H,d},\theta_{H,d})S^{(2)}_{H}(n_{H,d},\theta_{H,d})^{T}
\label{eqn:schematic}
\end{eqnarray}
where $S^{(2)}_{H}$ is defined by the Hessian kernel in Eq.(\ref{eqn:highoneloop}), the symbol $\int$ indicates integration over $\tau, \tau ' ,x ,x'$, and $\theta_{H,d}$ and $n_{H,d}$ are the quantum fluctuation fields. The partition function at this order contains a sum over all solutions $n_{H,0}$, $\theta_{H,0}$, $n_{L,0}$, $\theta_{L,0}$ of Eqs.(\ref{eqn:euler}), (\ref{eqn:statphase}), (\ref{eqn:statphaseadj}). In what remains of the derivation, we restrict to those phase configurations that satisfy the lowest-energy condition ${2\theta_{H,0} -2\theta_{L,0} \over \hbar} = (2k+1)\pi$. This restriction is valid throughout $\Omega$ for temperatures lower than the maximal energy scale associated with background phase differences in the $J_{L}$ and $J_{H}$ sector, i.e., for $k_{B}T\ll V_{0}\min_{x\in \Omega}n_{H,0}n_{L,0}$.   With the low-energy restriction now assumed, the ACS arising at one-loop order for the phase fluctuation field $\theta_{H,d}$ can be derived following the recipe in Appendix A.  The only difference occurs in the last term of the expression for ${\delta^{2} S \over \delta \theta_{H}(x,\tau)\delta \theta_{H}(x',\tau')}$ in Eq.(\ref{eqn:highoneloop}), which gives rise to a mass term for the field $\theta_{H,d}$ propagating on the ACS. The action becomes that of a Klein-Gordon boson with spacetime-dependent mass propagating on ACS:
\begin{eqnarray}
S&=&S_{L,0} + S_{H}[n_{H,0},\theta_{H,0};n_{L,0},\theta_{L,0}] \nonumber \\ &+&{1\over 2}
\int \sqrt{-g} \left[ \vphantom{{A\over B}}
g^{\mu \nu}\del_{\mu}\theta_{H,d}\del_{\nu}\theta_{H,d} \right. 
\nonumber \\ 
&+&\left. {4V_{0}n_{H,0}n_{L,0} \over \hbar^{2}\sqrt{-g}} \theta_{H,d}^{2}\right] .
\label{eqn:kgoneloop}
\end{eqnarray} 

In the following subsection, we go beyond one-loop order to derive the full effective theory on ACS for the field $\theta_{H,d}$ when the low energy phase matching condition ${2\theta_{H,0} -2\theta_{L,0} \over \hbar} = (2k+1)\pi$ is satisfied throughout the domain.

\subsection{Sine-Gordon interaction in the $J_{H}$ sector \label{sec:jhhighloop}}
From Eq.(\ref{eqn:polarform}), it is clear that the tree-level energy $S_{H}[n_{H,0},\theta_{H,0};n_{L,0},\theta_{L,0}]$ is minimized by the phase matching condition $\theta_{H,0} - \theta_{L,0} = (2k+1)\pi \hbar / 2$, $k\in \mathbb{Z}$. 
The sine-Gordon term in the effective action for $\theta_{H,d}$ is derived by summation of all higher-loop contributions $\delta^{n}S_{H} / \delta \theta_{H}^{n}$. Specifically, when the phase matching condition is satisfied for all $x\in \Omega$, the higher-loop contribution is given by \begin{multline}
 \sum_{n=1}^{\infty}\int{1\over 2n!}{\delta^{2n}S_{H} \over \delta \theta_{H}^{2n}}\Big\vert_{\theta_{H,0}}\theta_{H,d}^{2n} \\
 = \int V_{0}n_{H,0}n_{L,0}\left(1-\cos {2\theta_{H,d}\over \hbar}\right).
\label{eqn:higherloop}
\end{multline} 
In Eq.(\ref{eqn:higherloop}), the integral on the left hand side (right hand side) symbolizes $2n$ integrations over imaginary time variables and over the space $\Omega$ (symbolizes a single integration over imaginary time and space $\Omega$). Furthermore, the term in $\delta^{2}S_{H} / \delta \theta_{H}^{2}$ that contributes only to the ACS has been omitted. 

In addition to the standard approximations presented in Appendix A and the low-energy phase matching condition derived in Sec. \ref{sec:jhoneloop}, there is one more approximation that should be made which guarantees that the dynamics of the phase fluctuation $\theta_{H,d}$ is given by the sine-Gordon model on ACS. From Eq.(\ref{eqn:highoneloop}), one can see that there are additional contributions which can be summed exactly coming from the mixed functional derivatives:
\begin{widetext}
\begin{eqnarray}
\sum_{n=1}^{\infty}\int{1\over 2n+1 !}\left({\delta^{2n+1}S_{H} \over \delta n_{H}\delta \theta_{H}^{2n}} + {\delta^{2n+1}S_{H} \over \delta \theta_{H}^{2n}\delta n_{H}}\right)\Big\vert_{n_{H,0}}n_{H,d}\theta_{H,d}^{2n} &=& 2V_{0}\int_{[0,\beta \hbar]}\int_{\Omega}{1\over 3! \hbar^{2}}n_{L,0}n_{H,d}\theta^{2}_{H,d} -{1\over 5! \hbar^{4}}n_{L,0}n_{H,d}\theta^{4}_{H,d}+\ldots \nonumber \\ &=& \int_{[0,\beta \hbar]}\int_{\Omega}2V_{0}n_{L,0}n_{H,d}\left(1-{\hbar \sin \left( 2\theta_{H,d}/\hbar \right) \over 2\theta_{H,d}}\right)
\label{eqn:sineterm}
\end{eqnarray}
\end{widetext} 
when $\theta_{H,0} - \theta_{L,0} = (2k+1)\pi \hbar / 2$.  In the following, we omit this term from the analysis due to the fact that after Gaussian integration over the amplitude fluctuation field $n_{H,d}$, the function $1-\hbar\sin (2\theta_{H,d}/\hbar)/2\theta_{H,d}$ appears in the following two types of expressions: 1) in a term with characteristic energy scaling as $\mathcal{O}(V_{0}^{2})$ which can be neglected due to the weakness of the interaction, and 2)  in terms of the form $V_{0}n_{L,0}\left( \del_{j} \theta_{H,d} \right) \left(1-{\hbar \sin \left( 2\theta_{H,d}/\hbar \right) / 2\theta_{H,d}}\right)$, $j=0,1,2,3$, which can be assumed to approximately vanish if $\theta_{H,d}(\omega_{-n},-k)\approx \theta_{H,d}(\omega_{n},k)$ in Matsubara and momentum space. We also note that Eq.(\ref{eqn:sineterm}) vanishes as $\theta_{H,d} \rightarrow 0$, the same limit for which the sine-Gordon dynamics on ACS is well approximated by Klein-Gordon dynamics of a massive boson on ACS.

Taking this additional approximation into account, implementing the phase matching condition, and following the derivation of Appendix A for the ACS arising at one-loop order gives the action for the phase field fluctuations $\theta_{H,d}$, differing from the free action of a massless particle on ACS by the addition of a nonperturbative sine-Gordon interaction arising from the summation of higher-loop contributions shown in Eq.(\ref{eqn:higherloop}):
\begin{eqnarray}
S_{\text{sG}}&:=&{1\over 2}
\int \sqrt{-g} \left[ \vphantom{{A\over B}}
g^{\mu \nu}\del_{\mu}\theta_{H,d}\del_{\nu}\theta_{H,d} \right. 
\nonumber \\ 
&+&\left. {2V_{0}n_{H,0}n_{L,0} \over \sqrt{-g}}\cos(2\theta_{H,d}/\hbar) \vphantom{{A\over B}}\right] .
\label{eqn:sgacs}
\end{eqnarray}
In Eq.(\ref{eqn:sgacs}), the integral is over the imaginary time interval $[0,\beta \hbar]$ and over the space $\Omega$. 
We emphasize  that the above action is exact  
when the well-defined approximations of the present section and those of Appendix A hold. As is the case for the Klein-Gordon mass arising at one-loop order in Eq.(\ref{eqn:kgoneloop}),  the sine-Gordon mass exhibits a spacetime dependence. Using $\sqrt{-g}=n_{H,0}^{2}/m^{2}c_{s}$, with $c_{s}:=\left( V_{0}n_{H,0} / m \right)^{1/2}$ the local speed of sound in the $J_{H}$ sector (see Appendix A), the sine-Gordon mass is seen to be $2(V_{0}m)^{3/2}n_{L,0}/n_{H,0}^{1/2}$.   

The contravariant metric $g^{\mu\nu}$ [see Eq.(\ref{eqn:contravar})] depends on the gauge field $v$ and a solution pair $n_{H,0}$, $\theta_{H,0}$ of Eq.(\ref{eqn:euler}). It follows from the phase matching condition $\theta_{H,0} - \theta_{L,0} = (2k+1)\pi \hbar / 2$, $k\in \mathbb{Z}$, for the background phase fields that $\nabla \theta_{H,0} = \nabla \theta_{L,0}$. This low-energy configuration also implies that if one exchanges $L$ and $H$ throughout the above calculation, the phase fluctuations $\theta_{L,d}$ in the $J_{L}$ sector propagate on a space $g^{\mu\nu}$ with the same form as derived in this section. Deviation from the phase-matching condition has two consequences: 1) the $\theta_{H,d} \rightarrow -\theta_{H,d}$ symmetry of the action is broken, thereby resulting in nonconservation of the particle number in the $J_{L}$ and $J_{H}$ sectors, and, 2) the fluctuations in the $J_{L}$ and $J_{H}$ sectors propagate on different ACS geometries when their complementary sectors are, respectively, pinned to tree-level. Note that when the $\theta_{H,d}$ dynamics are described by Eq.(\ref{eqn:sgacs}), the effective sine-Gordon theory is locally destroyed when $n_{L,0}=0$ (in such regions, the $\theta_{H,d}$ field becomes a free massless particle on $g^{\mu \nu}$ as in Appendix A), or when the high-energy configuration $2(\theta_{H,0} - \theta_{L,0}) = 2k \pi \hbar $, $k\in \mathbb{Z}$, is generated.

\subsection{Sine-Gordon equation on ACS\label{sec:sgeqonacs}}

The equation of motion for $\theta_{H,d}$ resulting from taking $\delta S_{\text{sG}}/\delta \theta_{H,d} = 0$ in Eq.(\ref{eqn:sgacs}) has the form of a nonlinear wave equation on the ACS. Using Eq.(\ref{eqn:sgacs}) and setting the functional derivative $\delta S_{\text{sG}} / \delta \theta_{H,d} = 0$ gives the equation of motion
\begin{equation}
\del_{\mu}\left( \sqrt{-g}g^{\mu\nu}\del_{\nu}\theta_{H,d} \right) + {2V_{0}n_{H,0}n_{L,0} \over \hbar}\sin\left({2\theta_{H,d} \over \hbar} \right) = 0
\label{eqn:sgeoncurvedspacetime}
\end{equation}
in coordinates $(x_{0},x_{1},x_{2},x_{3}) = (-i\tau , x_{1},x_{2},x_{3} )$. In terms of the quantum theory of phase fluctuations in the $J_{H}$ sector, Eq.(\ref{eqn:sgeoncurvedspacetime}) is the sine-Gordon equation on curved spacetime that is satisfied by $\langle \hat{\theta}_{H,d} \rangle$ at tree order.
Written in real time, this equation is
\begin{eqnarray}
{}&{}&\del_{tt}\theta_{H,d} - \del_{t}\left( (v-{1\over m}\nabla \theta_{H,0}) \cdot \nabla \theta_{H,d} \right) \nonumber \\
&{}& - \nabla \cdot \left((v-{1\over m}\nabla \theta_{H,0}) \del_{t} \theta_{H,d} \right)\nonumber \\
&{}& -\nabla \cdot \left( \left({V_{0}n_{H,0}\over m}\mathbb{I}_{3\times 3} \nonumber \right. \right. \\ &{}& \left. \left. - (v-{1\over m}\nabla \theta_{H,0})^{T}(v-{1\over m}\nabla \theta_{H,0}) \right)\nabla\theta_{H,d} \right) \nonumber \\ &{}& + {2V_{0}^{2}n_{H,0}n_{L,0} \over \hbar} \sin\left( {2\theta_{H,d}\over \hbar} \right) = 0.
\label{eqn:onedsg}
\end{eqnarray}
When restricted to one spatial dimension, the above equation does not immediately reduce to the usual sine-Gordon equation $\left(\del_{t}^{2}-\del_{x}^{2}\right)\Phi + {M^{2}\over \beta}\sin (\beta \Phi)$ for a scalar field $\Phi(x,t)$ and constants $M$, $\beta >0$. Rather, the assumptions $\del_{x}v=0$, $\del_{xx}\theta_{H,0}=0$ that were used in our derivation of $S_{\text{sG}}$ imply that
\begin{eqnarray}
{}&{}&\del_{tt}\theta_{H,d}- 2\left(v-{1\over m}\del_{x}\theta_{H,0} \right) \del_{xt}\theta_{H,d} \nonumber \\ &{}&
-\left({V_{0}n_{H,0}\over m} - (v-{1\over m}\del_{x} \theta_{H,0})^{2} \right)\del_{xx} \theta_{H,0} \nonumber \\ &{}& +\del_{x}\left( {1\over m}\del_{t}\theta_{H,0} - {V_{0}n_{H,0}\over m} \right) \del_{x}\theta_{H,d} \nonumber \\ &{}&+ {2V_{0}^{2}n_{H,0}n_{L,0} \over \hbar} \sin\left( {2\theta_{H,d}\over \hbar} \right) = 0.
\label{eqn:odsg}
\end{eqnarray}
The assumption $\del_{xx}\theta_{H,0}=0$ means that $\del_{x}\theta_{H,0}$ is a function of time only. It follows from the stationary phase equation $\delta S_{H}/\delta n_{H}=0$ in Eq.(\ref{eqn:euler}) that if $\del_{x}\theta_{H,0}/m = v$ and if $n_{H,0}$ is well approximated by the Thomas-Fermi limit $n_{H,0}={1\over 2V_{0}}\left( \mu - V_{\text{ext}}(x) - V_{0}n_{L,0} \right)$, then  
\begin{equation}\del_{t}\theta_{H,0}= V_{0}n_{H,0}(x,t) \label{eqn:cond1}\end{equation} 
and, therefore, the term of Eq.(\ref{eqn:odsg}) linear in $\del_{x}\theta_{H,d}$ vanishes.
In this case, Eq.(\ref{eqn:odsg}) becomes
\begin{equation}
\left( \del_{t}^{2}-{V_{0}n_{H,0}\over m}\del_{x}^{2} \right)\theta_{H,d} + {2V_{0}^{2}n_{H,0}n_{L,0} \over \hbar}\sin \left( {2\theta_{H,d}\over \hbar} \right) = 0
\label{eqn:nonautosge}
\end{equation}
which, for nonconstant $n_{H,0}$ or $n_{L,0}$, is a nonautonomous partial differential equation. Equation (\ref{eqn:nonautosge}) can be put into the usual sine-Gordon form with $\beta = \hbar/2$ and $M^{2} = V_{0}^{2}n_{H,0}n_{L,0}$ by taking $n_{H,0}$, $n_{L,0}$ to be constant and changing from the laboratory coordinates $(\tau ,x)$ to the canonical coordinates \cite{evanspde}
for the second-order PDE.   In (1+1)-D, the gauge field $v$ is constant and can be completely removed from the equation of motion for $\theta_{H,d}$ by using Eq.(\ref{eqn:euler}). Specifically, when the background phases are in their lowest energy configuration, one has $ \left( v-{1\over m}\del_{x}\theta_{H,0}\right) = \del_{t}n_{H,0} / \del_{x}n_{H,0}$.

\section{Examples}
 
\subsection{Engineered condensate state\label{sec:engineer}}
Without loss of generality, the set $J$ in the previous section can be taken as countable, partially ordered and $J_{L}$ ($J_{H}$) considered as low-energy (high-energy) modes. In this subsection, we consider the effective action for a phase fluctuation field having support only on modes $J_{H}=\lbrace j >0 \rbrace$ when the $J_{L}=\lbrace 0 \rbrace$ mode is prepared in an engineered state. For example, one can take $\alpha \in \mathbb{C}$ and construct the subspace $\mathcal{B}_{\alpha}$ of the bosonic Fock space $\mathcal{F}_{B}$ defined as the completion of the complex linear span of pure states of the form
\begin{equation}
\ket{\psi_{\vec{n}}}:=e^{-{\vert \alpha \vert^{2}\over 2}}\sum_{j_{0}=0}^{\infty}{\alpha^{j_{0}}\over \sqrt{j_{0}!}}\ket{j_{0},n_{1},n_{2},\ldots }   
\end{equation}
where $\vec{n}:=(n_{1},n_{2},\ldots )$ with $n_{k}\ge 0$. Clearly, $\langle \psi_{\vec{n}'} \vert \psi_{\vec{n}} \rangle = \delta_{\vec{n},\vec{n}'}$ and $a_{0}\ket{\psi_{\vec{n}}}=\alpha \ket{\psi_{\vec{n}}}$ for all $\vec{n}$. We now compress the Hamiltonian Eq.(\ref{eqn:wibgham}) to the subspace $\mathcal{B}_{\alpha}$ by defining $\hat{H}_{\alpha}:= P_{\mathcal{B}_{\alpha}}\hat{H}P_{\mathcal{B}_{\alpha}}$. Explicitly, $P_{\mathcal{B}_{\alpha}}\hat{H}P_{\mathcal{B}_{\alpha}}$ is given by taking $\hat{\psi} \rightarrow \alpha \varphi_{0}(x) + \hat{\psi}_{H}$ in Eq.(\ref{eqn:wibgham}), where $\hat{\psi}_{H}$ has support only on $J_{H}$. Utilizing the compressed Hamiltonian $\hat{H}_{\alpha}$ to predict the thermodynamic properties of the WIBG is traditionally known as the Bogoliubov approximation \cite{bogo,zagrebnov,liebcnumber}. Within the Bogoliubov approximation, the partition function is written \begin{eqnarray} Z(\beta) &\approx& \text{tr}'\left[e^{-\beta P_{\mathcal{B}_{\alpha}}\hat{H}P_{\mathcal{B}_{\alpha}}}\right]\end{eqnarray} where $\text{tr}'$ is the trace over $\mathcal{B}_{\alpha}$ only and $\beta$ represents the inverse temperature as measured in subspace $\mathcal{B}_{\alpha}$. The field operator $\hat{\psi}$ has nonzero expectation value for any state of $\mathcal{B}_{\alpha}$, so we are working in the Bose-Einstein condensed phase.  We can also take $\alpha \in \mathbb{R}$ by making the change of variable $\hat{\psi}_{H} \mapsto e^{i\text{Arg}{\alpha}}\hat{\psi}_{H}$.

The approximate partition function can be written as a coherent state functional integral over functions orthogonal (in $L^{2}(\Omega)$) to $\varphi_{0}(x)$. When terms of order $\mathcal{O}(\psi_{H}^{3})$ and $\mathcal{O}(\psi_{H}^{4})$ are neglected in the action of this partition function, the thermodynamics of the $j>0$ sector is determined by a noninteracting gas of bosons with Bogoliubov spectrum \cite{abrikosov}. We now show that keeping these terms allows for derivation of sine-Gordon dynamics on ACS in the $j>0$ sector as in Sec. \ref{sec:sgdyn}. Similar to the procedure in Sec. \ref{sec:sgdyn}, we demand that the single-particle wave function $\varphi_{0}$ satisfy the time-independent generalized Gross-Pitaevskii equation given by an equation similar to Eq.(\ref{eqn:statphase}): 
\begin{eqnarray}
-{\hbar^{2}\over 2m}(\nabla -i{m\over \hbar}v(x))^{2}\varphi_{0} &+& (mV_{\text{ext}}(x)+ V_{0}\vert \psi_{H} \vert^{2})\varphi_{0} \nonumber \\ &+&  \vert \alpha\vert^{2}V_{0}\vert \varphi_{0} \vert^{2}\varphi_{0} = 0.
\label{eqn:gpusual}
\end{eqnarray}
Substituting a solution of the form $\varphi_{0} = \vert \varphi_{0} \vert e^{i\theta_{0} /\hbar}$ back into the action gives Eq.(\ref{eqn:polarform}) with $n_{L,0} \mapsto \vert \alpha \vert^{2}\vert \varphi_{0} \vert^{2}$ and $\theta_{L,0} \mapsto \theta_{0}$.

Because the resulting action takes the same form as Eq.(\ref{eqn:polarform}), therefore the phase fluctuation field $\theta_{H,d}$ with support only on single-particle modes $j>0$ exhibits sine-Gordon dynamics on ACS under the same assumptions as in Sec. \ref{sec:sgdyn}. Although we do not touch on the subject of spectral analysis of the effective field theory, we note that when the sine-Gordon interaction is neglected, the values of $\mu$ and $\alpha$ should be chosen so that ${\mu \over \vert \alpha \vert^{2} }= V_{0}$ to enforce gapless excitations \cite{griffinshi}.

One inconvenient feature of the derivation of the sine-Gordon dynamics on ACS in the $J_{H}$ sector presented in Sec. \ref{sec:sgdyn} is that the field with support $J\setminus J_{H}$ must satisfy the self-consistent mean field equations Eq.(\ref{eqn:statphase}) and Eq.(\ref{eqn:statphaseadj}). This requirement can be removed if the full dynamics is restricted to a subspace of $\mathcal{F}_{B}$ such that terms of the form $\hat{\psi}_{L}^{\dagger}\hat{\psi}_{H} +h.c.$, $\hat{\psi}_{L}^{\dagger}\hat{\psi}_{L}\hat{\psi}_{H} + h.c.$, and $\hat{\psi}_{H}^{\dagger}\hat{\psi}_{H}\hat{\psi}_{L} + h.c.$ vanish. As an example, consider again $J_{L}=\lbrace 0\rbrace$ and, instead of making the Bogoliubov approximation by projecting the Hamiltonian to the $j=0$ mode coherent state subspace $\mathcal{B}_{\alpha}$, prepare the $j=0$ mode so that the system dynamics occurs in the subspace $\mathcal{B}_{\alpha,+}\subset \mathcal{F}_{B}$, where $\mathcal{B}_{\alpha,+} $ is defined as the completion of the complex linear span of pure states of the form:
\begin{equation}
\ket{\psi_{\vec{n}}^{+}}:={1\over \sqrt{2+ 2e^{-2\vert \alpha \vert^{2}}}}\sum_{j_{0}=0}^{\infty}{\alpha^{j_{0}}(1+(-1)^{j_{0}})\over \sqrt{j_{0}!}}\ket{j_{0},n_{1},n_{2},\ldots }.
\label{eqn:evencat}
\end{equation}

One finds that  $\langle \psi_{\vec{n}}^{+} \vert a_{0}\vert{\psi_{\vec{n}'}^{+}} \rangle=0$ while $ a_{0}^{2}\vert{\psi_{\vec{n}}^{+}} \rangle=\alpha^{2}\vert{\psi_{\vec{n}}^{+}} \rangle$. The parameter $\alpha$ controls the expected number of atoms in the $j=0$ mode via $\langle \psi_{\vec{n}}^{+} \vert a_{0}^{\dagger}a_{0}\vert{\psi_{\vec{n}}^{+}} \rangle = \vert \alpha \vert^{2}\tanh^{2}\vert \alpha \vert^{2}$ and may be taken as real because of the global U(1) symmetry. Taking the partial trace of $\ket{\psi_{\vec{n}}^{+}}\bra{\psi_{\vec{n}}^{+}}$ over the bosonic Fock space generated from $\lbrace \varphi_{j}\rbrace_{j\in J\setminus \lbrace 0 \rbrace}$ gives the single-mode even coherent state 
(a photonic cat state), commonly studied in continuous variable quantum information theory \cite{volkoff,dodonov,milburncatcode}.

Compressed to the subspace $\mathcal{B}_{\alpha,+}$, the normally ordered Hamiltonian becomes
\begin{eqnarray}
P_{\mathcal{B}_{\alpha,+}}:\hat{H}:P_{\mathcal{B}_{\alpha,+}} &=& \left[\vphantom{\sum_{k}^{\infty}}  H[\overline{\alpha}\overline{\varphi_{0}},\alpha \varphi_{0}] \nonumber \right. \\ & +& \left.\int_{\Omega} \left[\vphantom{{V_{0}\over 2}}  {\hbar^{2}\over 2m}\overline{D} \hat{\psi}_{H}^{\dagger}D\hat{\psi}_{H} \nonumber \right. \right. \\ & +& \left.  \left.
(mV_{\text{ext}}(x) - \mu)\hat{\psi}_{H}^{\dagger}\hat{\psi}_{H} 
\nonumber \right. \right. \\ &+ & \left. \left.
2V_{0}\vert \varphi_{0} \vert^{2}\vert \alpha \vert^{2}\tanh^{2}\vert \alpha \vert^{2}\hat{\psi}_{H}^{\dagger}\hat{\psi}_{H} \nonumber  \right. \right. \\ & +& \left.  \left.
\left({V_{0}\over 2}\alpha^{2}\varphi_{0}^{2}\hat{\psi}_{H}^{\dagger \,2} + h.c. \right) \nonumber \right. \right. \\ & +& \left. \left.
{V_{0}\over 2}\hat{\psi}_{H}^{\dagger \, 2}\hat{\psi}_{H}^{2} \vphantom{{V_{0}\over 2}}\right] \vphantom{\sum_{k}^{\infty}} \right]P_{\mathcal{B}_{\alpha,+}}
\label{eqn:catham}
\end{eqnarray}
where $H[\overline{\alpha}\overline{\varphi_{0}},\alpha \varphi_{0}]$ is a scalar obtained by the formal substitution $\hat{\psi} \rightarrow \alpha \varphi_{0}$, $\hat{\psi}^{\dagger}\rightarrow \overline{\alpha}\overline{\varphi_{0}}$ in Eq.(\ref{eqn:wibgham}) and $h.c.$ symbolizes the adjoint of the preceding term in the parentheses. Compared to Eq.(\ref{eqn:actionfull}), Eq.(\ref{eqn:catham}) does not contain the terms $ :a_{0}^{\dagger}a_{0}a_{0}^{\dagger}\hat{\psi}_{H}: +h.c.$ or $:a_{0}\hat{\psi}_{H}^{\dagger}\hat{\psi}_{H}\hat{\psi}_{H}^{\dagger}: +h.c.$ because single-particle tunneling events involving the $j=0$ mode are forbidden in $\mathcal{B}_{\alpha,+}$. As a consequence of this partial decoupling, $\varphi_{0}$ is no longer required to satisfy a generalized, self-consistent Gross-Pitaevskii equation.
The phase-fluctuation dynamics in the $J_{H}$ sector are given by a sine-Gordon potential on ACS because the $J_{H}$ sector of the coherent state path integral is equivalent to Eq.(\ref{eqn:polarform}). In both the zero-mode coherent state case and the zero-mode even coherent state case, the sine-Gordon potential is given by $2V_{0}\vert \alpha \vert^{2}\vert \varphi_{0} \vert^{2}\left( 1- \cos\left( {2\theta_{H,d}\over \hbar} \right) \right)$ when the background phases are in the lowest energy configuration.

Any single-particle wave function orthogonal to the $J_{H}$ sector and prepared in an even coherent state can support sine-Gordon dynamics on analogue curved spacetime for the phase fluctuations in $J_{H}$. The statistical thermodynamics of the $J_{H}$ ``universe'' is determined by $P_{\mathcal{B}_{\alpha,+}}\left(:\hat{H}:-H[\overline{\alpha}\overline{\varphi_{0}},\alpha \varphi_{0}]\right) P_{\mathcal{B}_{\alpha,+}}$ and the single-particle wave function chosen for $\varphi_{0}$ appears only in setting the spacetime geometry, mass, and vacuum energy for the phase fluctuations.

\subsection{Interplane tunneling between WIBGs\label{sec:interplane}}

Another experimentally relevant class of sine-Gordon dynamics on ACS arises when both $J_{L}$ and $J_{H}$ sectors exhibit phase fluctuations, but the sectors are not coupled by $s$-wave WIBG scattering. An example of this situation is encountered in a system consisting of two planar (i.e., (2+1)-D) reservoirs of WIBG exhibiting single-particle tunneling with amplitude $t_{\perp}$ independent of time $\tau$ and the planar coordinate $x \in \Omega$:
\begin{eqnarray}
S_{T}&=&{t_{\perp}\over 2}\int_{[0,\beta\hbar]}\int_{\Omega}  \overline{\psi}_{L}\psi_{R} + c.c.
\end{eqnarray}
In this model, the phase fluctuations in each plane are coupled only by the interplane tunneling. Appealing to the general derivation in Appendix A, one finds that for $t_{\perp}\rightarrow 0$ the WIBG part of the action describes two independent massless boson fields $\theta_{L,d}$, $\theta_{R,d}$ propagating on ACS defined by the velocity fields $v_{L}+{1\over m}\nabla \theta_{L,0}$, $v_{R}+{1\over m}\nabla \theta_{R,0}$, respectively. For $\vert t_{\perp} \vert \neq 0$, however, the higher-order functional derivatives are
\begin{eqnarray}
{1\over 2n!}{\delta^{2n}S_{LR}\over \delta \theta_{L}^{m}\delta \theta_{R}^{k} }&=& {(-1)^{n+k}t_{\perp}\sqrt{n_{L}n_{R}}\over 2n!\hbar^{2n}}\cos \left( {\theta_{L} - \theta_{R} \over \hbar} \right), \nonumber \\ 
{1\over 2n+1!}{\delta^{2n+1}S_{LR}\over \delta \theta_{L}^{m}\delta \theta_{R}^{k} }&=& {(-1)^{n+k+1}t_{\perp}\sqrt{n_{L}n_{R}}\over 2n+1!\hbar^{2n+1}}\sin \left( {\theta_{L} - \theta_{R} \over \hbar} \right)\nn 
\end{eqnarray}
where $m+k = 2n$ ($m+k = 2n+1$) in the first (second) line and in the $(n,m,k)=(1,2,0)$ and $(n,m,k)=(1,0,2)$ terms in the first line, we have neglected free Bose gas contribution to the second-order functional derivative. These higher loop contributions can be summed to produce the following action:
\begin{eqnarray}
S_{LR}&=& \int_{[0,\beta \hbar]}\int_{\Omega} \sum_{j\in \lbrace L , R \rbrace}\sqrt{-g_j}\left[\vphantom{\sum_{j\in \lbrace L , R \rbrace}}g_j^{\mu\nu}\del_{\mu}\theta_{j,d}\del_{\nu}\theta_{j,d} \right. \nonumber \\ &+& \left. t_{\perp}{\sqrt{n_{H,0}n_{L,0}}\over \sqrt{-g_j}} \left[ \cos \gamma_{LR,0}\left(1-\cos  \gamma_{LR,d} \right) \nonumber \right. \right. \\ &+& \left. \left. \sin  \gamma_{LR,0}  \left(\gamma_{LR,d}-\sin \gamma_{LR,d} \right) \right] \vphantom{\sum_{j\in \lbrace L , R \rbrace}} \right]
\label{eqn:leftrightplane}
\end{eqnarray}
where $\gamma_{LR}:= {\theta_{L}-\theta_{R} \over \hbar}$, the 0 subscript on the fields indicates that they are solutions to the coupled stationary phase equations for $\theta_{L(R)}$, $n_{L(R)}$, and the $d$ subscript indicates fluctuation fields. 

Equation (\ref{eqn:leftrightplane}) becomes the action of a Josephson tunnel junction between the $L$ and $R$ ``universes" in the low-energy configuration given by $\gamma_{LR,0}$ an odd multiple of $\pi$; a discussion of oscillations around $\gamma_{LR,0} = \pi$ in bosonic Josephson junctions can be found in \cite{Raghavan}. 
Otherwise, the junction has mixed sinusoidal phase dynamics. As in the previous sections, partitioning the single-particle Hilbert spaces of $L$ and $R$ systems would lead to internal Josephson oscillations in each system coupled to the external Josephson oscillation between the systems. 

\section{Quantum ACS from two-mode superposition state\label{sec:superposacs}}

The analysis in Sec. \ref{sec:engineer} suggests the possibility of projecting more than one mode to an engineered state of interest. In this section, we briefly analyze the dynamics defined by $P_{\mathcal{B}^{(2)}_{w,\alpha}}\hat{H}P_{\mathcal{B}^{(2)}_{w,\alpha}}$, where $B^{(2)}_{w,\alpha}\subset \mathcal{F}_{B}$ is defined as the completion of the complex linear span of pure states of the form
\begin{eqnarray}
\ket{\psi_{\vec{n}}}&:=& {e^{-\vert \alpha \vert^{2}/2}\over \sqrt{\mathcal{N}(w,\alpha)}}\left( \sum_{j_{0}=0}^{\infty}{\alpha^{j_{0}}(1+(-1)^{j_{0}})\over \sqrt{j_{0}!}}\ket{j_{0},0,n_{2},n_{3},\ldots }   \nonumber \right. \\ &+&\left. w \sum_{j_{1}=0}^{\infty}{\alpha^{j_{1}}(1+(-1)^{j_{1}})\over \sqrt{j_{1}!}}\ket{0,j_{1},n_{2},n_{3},\ldots } \right) 
\label{eqn:hierarchmany}
\end{eqnarray}
where $\mathcal{N}(w,\alpha):=4e^{-\vert \alpha \vert^{2} }\left( (1+\vert w \vert^{2})\cosh \vert \alpha \vert^{2} + 2\text{Re}w \right)$, $w\in \mathbb{C}$,  and $\vec{n} = (n_{2},n_{3},\ldots)$. Compression of the original Hamiltonian to this subspace allows for (1) taking $\varphi_{0}(x)$, $\varphi_{1}(x)$ to be arbitrary orthogonal single-particle wave functions in the $J_{L}$ sector and (2) dependence of the metric $g^{\mu\nu}$ on both background superfluid flows ${\hbar\over m}\nabla\text{Arg}\varphi_{0}$ and ${\hbar\over m}\nabla\text{Arg}\varphi_{1}$. Taking the partial trace of the state in Eq.(\ref{eqn:hierarchmany}) over the bosonic Fock space generated by single-particle modes $j>1$ gives a symmetrized two-mode state closely related to the hierarchical Schr\"{o}dinger cat states introduced in Ref.\cite{volkoffcat}.

Similarly to Eq.(\ref{eqn:catham}), compressing the Hamiltonian to the subspace $B^{(2)}_{w,\alpha}$ produces a Hamiltonian of the form (again using the global U(1) invariance to take $\alpha \in \mathbb{R}$)
\begin{eqnarray}
P_{\mathcal{B}^{(2)}_{w,\alpha}}\hat{H}P_{\mathcal{B}^{(2)}_{z,\alpha}} &=& \left[ \vphantom{\sum_{k}^{\infty}} \sum_{j\in \lbrace 0,1\rbrace}H[\overline{\varphi_{j}},\varphi_{j}] \nonumber \right. \\  &+&  \left. {2V_{0}\vert \alpha \vert^{4}e^{-\vert \alpha \vert^{2}} \over \mathcal{N}(\alpha, w)}\left( w\varphi_{1}^{2}\overline{\varphi_{0}}^{2} + c.c. \right) \nonumber \right. \\  &+&  \left. \int_{\Omega} \left[ \vphantom{{A\over B}} {\hbar^{2}\over 2m}\overline{D} \hat{\psi}_{H}^{\dagger}D\hat{\psi}_{H} \nonumber \right. \right. \\  &+&  \left. \left. (mV_{\text{ext}}(x) - \mu +G[\varphi_{0},\varphi_{1};\alpha ,w])\hat{\psi}^{\dagger}_{H}\hat{\psi}_{H}  \nonumber \right. \right. \\  &+&  \left. \left. {V_{0}\over 2}{\alpha^{2}\over \mathcal{N}(\alpha ,w)}\left((\overline{\varphi_{0}}^{2}+w^{2}\overline{\varphi_{1}}^{2})\hat{\psi}_{H}^{2}+h.c \right) \nonumber \right. \right. \\  &+&  \left. \left. {V_{0}\over 2}\hat{\psi}_{H}^{\dagger \, 2}\hat{\psi}_{H}^{2}    \vphantom{{A\over B}} \right] \vphantom{\sum_{k}^{\infty}} \right]P_{\mathcal{B}^{(2)}_{z,\alpha}}
\label{eqn:twomodefixed}
\end{eqnarray}
where the function $G[\varphi_{0},\varphi_{1};\alpha ,w]) := 2V_{0}\vert \alpha\vert^{2}\tanh\vert \alpha\vert^{2}\left( {\vert \phi_{0} \vert^{2} + \vert w \varphi_{1}\vert^{2} \over \mathcal{N}(w,\alpha)} \right)$, and $H[\overline{\varphi_{j}},\varphi_{j}]$ is a scalar functional which contributes only to the vacuum energy of the phase fluctuation theory. As $\alpha \rightarrow \infty$, the intermode interaction energy in the $J_{L}$ sector appearing in the second line of Eq.(\ref{eqn:twomodefixed}) vanishes. In the coherent state path integral expression for the partition function corresponding to $P_{\mathcal{B}^{(2)}_{w,\alpha}}\hat{H}P_{\mathcal{B}^{(2)}_{z,\alpha}}$, the pair exchange term appearing in the fifth line of Eq.(\ref{eqn:twomodefixed}) is given by $\vert \varphi_{0}^{2} + \vert w\vert ^{2}\varphi_{1}^{2}\vert {V_{0} \alpha^{2} \over \mathcal{N}(w,\alpha)}\cos({2\theta_{H} \over \hbar} - \xi )$ where $\xi := \text{Arg}\left( \varphi_{0}^{2} + \vert w\vert^{2} \varphi_{1}^{2} \right)$.

\begin{figure}[t]
\includegraphics[scale=0.33]{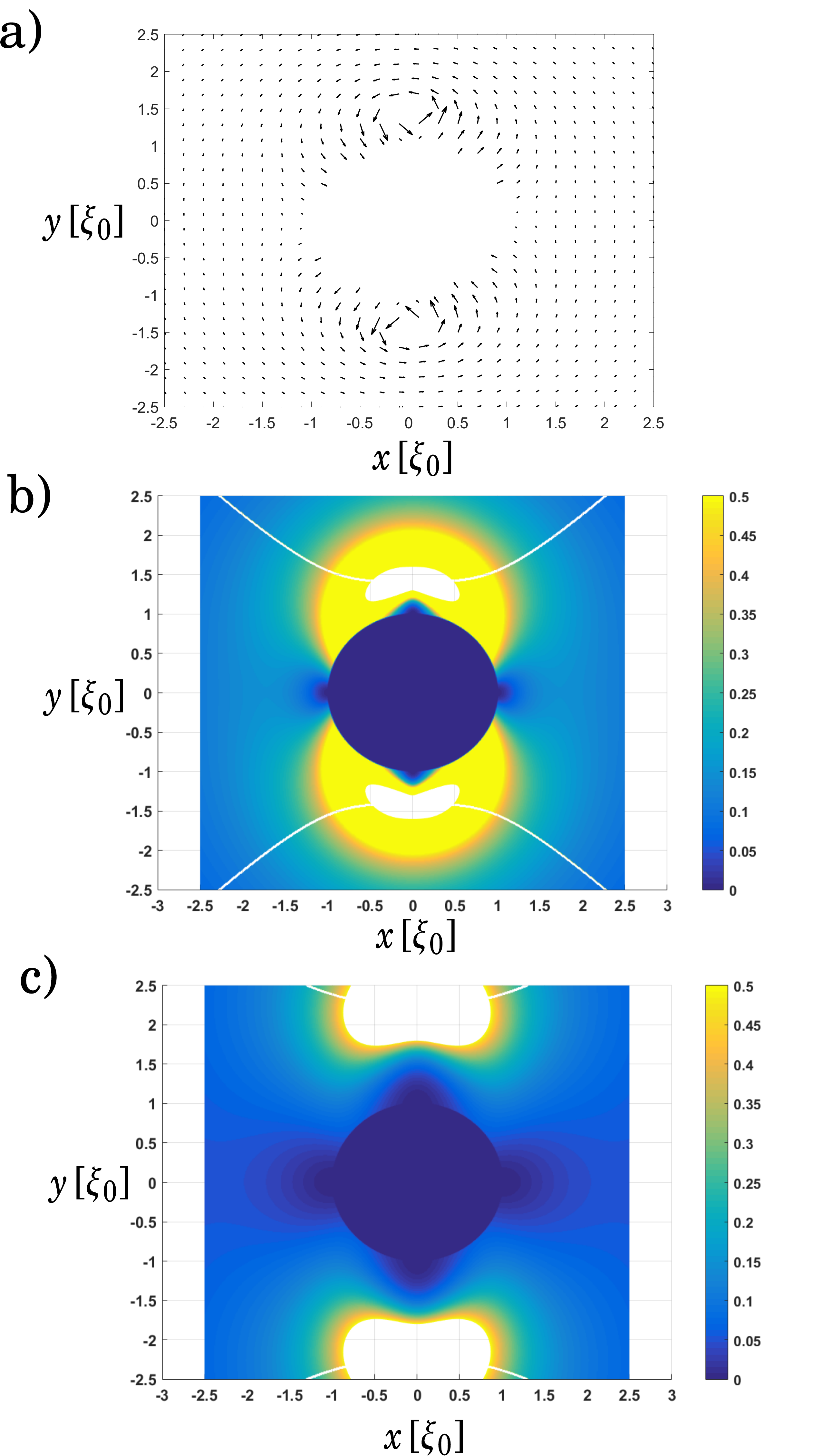}
\caption{a) Velocity field $\nabla \theta_{H,0}$ corresponding to $\theta_{H,0}$ in Eq.(\ref{eqn:vortnovort}) for $\Vert x \Vert \ge 1$ and $w=1$ in Eq.(\ref{eqn:hierarchmany}).  The gauge field $v$ is taken to be zero, and lengths are scaled by the 
superfluid coherence length $\xi_0$ appearing in \eqref{eqn:vortex}. b) Magnitude of velocity field when $w=1$ in Eq.(\ref{eqn:vortnovort}), with unit of velocity $2c_{s}$. The yellow region (representing velocity magnitude $\ge 1/2$ because the color scale has been cut off at 1/2) lies inside the ergosurface. The vortex core is given by the disk of radius 1; fluid velocity has been set to zero in the core. The white region is a neighborhood of a phase singularity surface. c) Same as b) except with $w=1/2$.
}
\label{fig:ergovel}
\end{figure}

Constructing the partition function via a trace of $\exp \left[ - \beta P_{\mathcal{B}^{(2)}_{w,\alpha}}\hat{H}P_{\mathcal{B}^{(2)}_{z,\alpha}}  \right]$ over $B^{(2)}_{w,\alpha}$ and proceeding in the same way as in previous sections, one can derive the sine-Gordon theory on ACS for the phase fluctuation field $\theta_{H,d}$. The intriguing novelty in the present case is that the low-energy tree-level configuration for $\theta_{H,0}$, which appears in $g^{\mu\nu}$ for the phase fluctuation field, is given by $\theta_{H,0} = \hbar \text{Arg}\left( \varphi_{0}^{2} + \vert w\vert^{2} \varphi_{1}^{2} \right) /2 + (2k+1)\hbar\pi /2$. Therefore, the metric contains contributions from both background velocity fields $\hbar\text{Arg} \nabla\varphi_{0}/m$ and $\hbar\text{Arg}\nabla \varphi_{1}/m$. Furthermore, the parameter $w$ that weights the $J_{L}$ sector superposition state toward being an even coherent state in the $j=0$ mode ($z\rightarrow 0$) or in the $j=1$ mode $z\rightarrow \pm\infty$ appears in the metric tensor, causing $\xi \rightarrow 2\text{Arg}\varphi_{0}$ ($w\rightarrow 0$) or $\xi \rightarrow 2\text{Arg}\varphi_{1}$ ($w\rightarrow \pm \infty$). In essence, the preparation of the state Eq.(\ref{eqn:hierarchmany}) is equivalent to the preparation of a false vacuum \cite{coleman} for the phase fluctuation field $\theta_{H,d}$. If a low energy atom is detected, the superposition collapses to either the $\varphi_{0}$ even coherent state or the $\varphi_{1}$ even coherent state conditioned on the low energy atom being found in $\varphi_{0}$ or $\varphi_{1}$, respectively. Because the locally-defined coupling constant of the sine-Gordon theory for $\theta_{H,d}$ is proportional to $\cos(2\theta_{H,0}/\hbar - \xi)$, the collapse event changes the energy density of the system. 

Finally, we mention that in the present case, the properties of the ACS geometry can arise from classically disallowed background flows. Consider a (2+1)-D example when $\varphi_{0}$ is a U(1) quantum vortex centered on the origin with circulation $2\pi \hbar n / m$. The following unnormalized approximate wave function is valid for $\Vert x\Vert > \xi_{0}$ \cite{pitaevskii}
\begin{equation}
\varphi_{0}(x)
=A e^{in\tan^{-1}\left({x_{2}\over x_{1}}\right)}\left(1-{\xi_{0}^{2}n^{2}\over \Vert x \Vert^{2}}\right)^{1/2}
\label{eqn:vortex}
\end{equation}
with $A$ a positive constant and $\xi_{0}$ a microscopic length scale characterizing the radius of the vortex core (the superfluid coherence length). The superposed single-particle mode $\varphi_{1}$ can be taken as a U(1) quantum vortex of the same form as above and also centered at the origin, but with different circulation $2\pi \hbar n'/m$. The quantum nature of such a superposition can be seen by noting that there exists an annular region in the gas which is in a superposition of the gas state comprising the vortex core and the superfluid state.

From $g_{00}$ in Eq.(\ref{eqn:covar}), one can determine the ergosurface condition in terms of $\Vert v-{1\over m}\nabla \theta_{H,0}\Vert^{2}$. Taking $\theta_{H,0}$ to be unitless for now and taking $v=0$ gives the condition ${\hbar^{2}\over m^{2}}\Vert \nabla \theta_{H,0}\Vert^{2} = c_{s}^{2}$. Working in units where $\xi_{0} = 1$, the ergosurface is given by $(x,y)$ such that $\Vert \nabla \theta_{H,0}(x,y)\Vert^{2} = 1/2$. In the absence of a gauge field, the connection coefficients, Riemann curvature tensor, Ricci curvature tensor, scalar curvature, and Einstein tensor for a U(1) vortex velocity field have been computed in Ref.\cite{fischervisser1,fischervisser2} in terms of the deformation rate $D_{ij}={1\over 2}(\del_{i}\del_{j} \theta_{H,0}+\del_{j}\del_{i} \theta_{H,0})$ when the $n_{H,0}$ and the speed of sound are taken to be spatially constant. Therefore, the Riemannian geometry of the spacetime is completely determined by the background velocity field. In Fig.\ref{fig:ergovel}, we show the velocity field and ergosurface for two values of $w$ in \eqref{eqn:hierarchmany}
when $\varphi_{0}$ is a vortex of circulation $n=1$ given by Eq.(\ref{eqn:vortex}), and $\varphi_{1}$ is taken as spatially constant. The superfluid velocity potential is
\begin{multline}
\theta_{H,0}(x_{1},x_{2}) \\
= {1 \over 2}\tan^{-1}\left( {2w^{2}(x^{2}_{1}+x_{2}^{2}-1)x_{1}x_{2} \over (1-w^{2})x_{1}^{2}+(1+w^{2})x_{2}^{2} + w^{2}(x_{1}^{4}-x_{2}^{4})}\right).
\label{eqn:vortnovort}
\end{multline}
From Fig.\ref{fig:ergovel}a), b), it is clear that the superposition of a spatially homogeneous state and an $n=1$ vortex in the lowest two modes drastically alters the velocity field from the azimuthal field of a U(1) quantum vortex. Although it appears that the U(1) symmetry has been broken, this is an artifact of having chosen $w,\alpha \in \mathbb{R}$. In reality, the phase singularities can occur along any direction, but engineering of the state of the $J_{L}$ sector can break the symmetry. When $w$ is decreased toward zero in Fig.\ref{fig:ergovel}c), the velocity field exhibits a larger ``quiet area'' separating the ergosurface from the vortex core. The regions of highest velocity are associated with phase singularities. For the phase field in Eq.(\ref{eqn:vortnovort}), the local coupling constant of the effective sine-Gordon theory depends only on $\vert \varphi_{0}^{2} + w^{2}\varphi_{1}^{2}\vert$ because the phase field $\theta_{H,0}$ is pinned to $\hbar \xi /2$ in Eq.(\ref{eqn:twomodefixed}). Therefore, the velocity fields shown in Fig.\ref{fig:ergovel} affect the dynamics of the phase fluctuation field $\theta_{H,d}$ only through their appearance in the analogue metric $g^{\mu\nu}$.

\section{Conclusion}

By expanding the action functional appearing in the coherent state path integral for the partition function of the WIBG, we have shown that the sine-Gordon model on ACS arises as the effective theory for phase fluctuations in the WIBG when the fluctuations are restricted to a subspace of the single-particle Hilbert space. From our analysis of the ACS existing on top of a low-energy mode prepared in a coherent state or even coherent state, one can see that the effective spacetime arising in the $J_{H}$ sector depends on how the sectors $J_{L}$ and $J_{H}$ comprising the bipartition are coupled.
We considered the case of coupled phase fluctuations in both mode sectors by analyzing a system consisting of tunnel-coupled (2+1)-D planes of WIBG. To demonstrate the dramatic effects of quantum state engineering on the effective spacetime for the high-energy phase fluctuations,  we calculated the ACS that arises when two low-energy single-particle modes are prepared in a macroscopic superposition state.

In this paper, we have not delved into methods for generating the states of the quantum vacuum (i.e., the states of $J_{L}$) on which the sine-Gordon model lives. One could envision a combination of optical pumping and stirring to tune the low energy particle occupation statistics and superfluid velocity profile, respectively. In any case, the combination of control of atomic transitions and hydrodynamic regimes makes ultracold alkali gases an ideal experimental setting for realization of the present dynamics. Our method for generating quantum sine-Gordon dynamics within well-defined physical constraints is expected to provide a platform for simulation of interacting (2+1)- and (3+1)-D quantum field theory on ACS. 

\acknowledgments 

This research was supported by the NRF Korea, Grant No. 2014R1A2A2A01006535.

\appendix*

\section{ACS IN FINITE-TEMPERATURE
WIBG WITHOUT MODE PARTITION\label{sec:app1}}
If the phase fluctuation field is allowed to have components on all single-particle wave functions, i.e., if it is not restricted to any sector, the phase fluctuation is a free, massless boson on ACS. A standard derivation of this fact is predicated on three conditions \cite{barcelorev,BLVBEC}: 1) making the Bogoliubov approximation for the field operator $\hat{\psi}$, 2) making a self-consistent mean field approximation to the Heisenberg equation of motion generated by $\hat{H}$ in Eq.(\ref{eqn:wibgham}) and neglecting particle number nonconserving products of field operators, and 3) neglecting the contribution of phase field fluctuations on a length scale smaller than the healing length of the WIBG. Here, we provide a derivation showing that propagation of a massless bosonic field on ACS arises simply as the one-loop contribution to the action of a locally gauge invariant WIBG. The derivation is predicated on three minimal physical assumptions, where we recall that $v(x)$ is the U(1) gauge field in Eq.(\ref{eqn:wibgham}) and $(n_{0}(x),\theta_{0}(x))$ is a solution pair to the coupled stationary phase equations for the amplitude and argument of the field $\psi$:
\begin{itemize}
\item Assumption 1: $\nabla \cdot v(x) = 0$. This assumption can be disposed of if there is no gauge field or external flow.
\item Assumption 2: $\nabla \cdot \nabla \theta_{0} = 0$. This assumption requires that the solution of the stationary phase equation be harmonic on $\Omega$.
\item Assumption 3: ${\hbar^{2} \over 8m}n^{-1}\nabla n \cdot \nabla n = 0$. This assumption is the same as the third condition above. It is assumed to hold as an identity for the positive semidefinite operator $\hat{\psi}^{\dagger}\hat{\psi}$, not just at the level of equations of motion.
\item Assumption 3': The two loop contribution $\mathcal{O}((\nabla\theta_{d})^2)$ vanishes, where $\theta_{d}$ is the phase fluctuation field introduced below. As with assumption 3, it is also a condition on a quantum field.
\end{itemize}

Assumption 1 is equivalent to assumption 2 precisely when the gauge field is exact, i.e., can be written as the gradient of some scalar. In that case, the gauge field can be canceled by an appropriate local U(1) gauge transformation on the fields $\psi$, $\overline{\psi}$ and assumption 2 becomes sufficient. Assumption 3' is related to assumption 3 in that both assumptions are satisfied if the theory is allowed to hold only on a length scale greater than the healing length $\xi_{0}$. In terms of our general treatment in terms of the single-particle Hilbert space spanned by orthonormal wave functions $\varphi_{j}$, we should assume that these do not vary greatly on length scales $\lesssim \xi_{0}$. In the coherent state path integral for the partition function associated with Eq.(\ref{eqn:wibgham}), one makes the change of field variables $\psi \mapsto \sqrt{n}e^{i\theta / \hbar}$. The stationary phase equations are given by
\begin{widetext}
\begin{eqnarray}
{\delta S \over \delta n} = 0 &\Rightarrow& i \del_{\tau}\theta + mV_{\text{ext}}(x,\tau)-\mu + {m\over 2}v\cdot v +{1\over 2m}\nabla \theta \cdot \nabla \theta -  v\cdot \nabla \theta + V_{0}n = 0 , \\
{\delta S \over \delta \theta} = 0 &\Rightarrow& -i \del_{\tau}n - {1\over m}\nabla \cdot (n\nabla \theta) +  \nabla \cdot (nv) = 0 \nonumber \\ &\Rightarrow& -i \del_{\tau}n +\nabla n \cdot \left( v- {1\over m}  \nabla \theta \right) +n \nabla \cdot \left( v - {1\over m}\nabla \theta \right) = 0 \label{eqn:statph1}\\ &\Rightarrow& -i\del_{\tau}n +\nabla n \cdot \left( v- {1\over m}  \nabla \theta \right) - {1\over m} n \nabla^{2}\theta = 0 
\label{eqn:statphase2} \\ &\Rightarrow & -i \del_{\tau}n +\nabla n \cdot \left( v- {1\over m} \nabla \theta \right) = 0 
\label{eqn:statphase3}
\end{eqnarray} 
where in passing from Eq.(\ref{eqn:statph1}) to Eq.(\ref{eqn:statphase2}), we have used assumption 1 and in passing from Eq.(\ref{eqn:statphase2}) to Eq.(\ref{eqn:statphase3}) we have used assumption 2 because $(\theta_{0},n_{0})$ is  defined to be a solution to these coupled equations.

The second-order functional derivatives of $S$ are:

\begin{eqnarray}
\delta^{2} S \over \delta n(x,\tau)\delta n(x',\tau') &=& V_{0}  \delta(x-x')\delta(\tau-\tau ') ,\nonumber \\
{\delta^{2} S \over \delta n(x,\tau)\delta \theta(x',\tau')} &=& -i \del_{\tau}\delta(x-x')\delta(\tau-\tau ') -{1\over m}\nabla \theta_{0}\cdot \nabla \delta(x-x')\delta(\tau-\tau ') - {1\over m}\delta(x-x')\delta(\tau-\tau ')\nabla^{2}\theta_{0} \nonumber \\ &{}& +  v\cdot \nabla \delta(x-x')\delta(\tau-\tau '), \nonumber \\ {\delta^{2} S \over \delta \theta(x,\tau)\delta n(x',\tau')}&=&i \del_{\tau}\delta(x-x')\delta(\tau-\tau ') + {1\over m} \nabla \theta_{0} \cdot \nabla \delta(x-x')\delta(\tau-\tau ') -  v\cdot \nabla \delta(x-x')\delta(\tau-\tau '), \nonumber \\ {\delta^{2} S \over \delta \theta(x,\tau)\delta \theta(x',\tau') } &=& -{1\over m}\nabla n_{0} \cdot \nabla \delta(x-x')\delta(\tau-\tau ') - {1\over m}n_{0}\nabla^{2}\delta(x-x')\delta(\tau-\tau ') .
\label{eqn:usualsecond}
\end{eqnarray}

Note that in first line in Eq.(\ref{eqn:usualsecond}), the condition $V_{0}\neq 0$ is necessary for the field $n_{H,d}$ to appear at quadratic order in the action and thereby promote the nonrelativistic dynamics of the bosonic field $\theta_{H,d}$ to dynamics on ACS. Introducing the fluctuation fields $n_{d}(x,\tau)$ and $\theta_{d}(x,\tau)$, the action expanded to one-loop order is given by $S=S^{(0)} +S^{(2)}$ where  $S^{(0)}$ is the tree-order action,
\begin{eqnarray}
S^{(2)} &=&{1\over 2}\int_{0}^{\beta \hbar}{d\tau \over \hbar}\int_{\Omega}d^{3}x \, \left[ V_{0}n_{d}^{2} + n_{d}\left( -2i\del_{\tau}\theta_{d} - {2\over m}\del_{j}\theta_{d}\del_{j}\theta_{0} + 2 v_{j}\del_{j}\theta_{d} \right) 
+ {1 \over m}n_{0}\del_{j}\theta_{d}\del_{j}\theta_{d} \right]
\end{eqnarray}
and where repeated summation over spatial index $j$ is implied. Gaussian integration over the real fluctuation $n_{d}$ field results in the following expression:
\begin{equation}
S^{(2)} = -{1\over 2V_{0}}\int_{0}^{\beta \hbar}{d\tau \over \hbar}\int_{\Omega}d^{3}x \, \left[ \left( -i\del_{\tau}\theta_{d} + \nabla\theta_{d}\cdot(v-{1\over m}\nabla \theta_{0}) \right)^{2} -V_{0}{1\over m}n_{0}\nabla\theta_{d}\cdot \nabla\theta_{d}\right].
\end{equation}

Taking $(x_{1},x_{2},x_{3})\in \Omega$ and $x_{0} = -i\tau$  so that $\del_{0}\theta_{d} = i\del_{\tau}\theta_{d}$ gives the action up to one-loop order
\begin{equation}
S=S[n_{0},\theta_{0}]+{1\over 2}\int_{0}^{\beta \hbar}{d\tau \over \hbar}\int_{\Omega}d^{3}x\, f^{\mu \nu}\del_{\mu}\theta_{d}\del_{\nu}\theta_{d}
\label{eqn:masslessact}
\end{equation}
where
\begin{eqnarray}
f^{00}=-{1\over V_{0}},\qquad 
f^{0j}={1\over V_{0}}\left( v_{j}(x)-{1\over m}\del_{j}\theta_{0}(x) \right),\qquad 
f^{j0}=f^{0j},  \nonumber \\
V_{0}f^{ij} =  {V_{0}\over m}n_{0}(x)\delta_{ij} -\left( v_{i}(x)-{1\over m}\del_{i}\theta_{0}(x) \right)\left( v_{j}(x)-{1\over m}\del_{j}\theta_{0}(x) \right) .
\end{eqnarray}

Equation (\ref{eqn:masslessact}) can be written in the canonical form of integration over a compact Riemannian manifold by finding $g^{\mu \nu}$ such that $\sqrt{-g}g^{\mu\nu}=f^{\mu\nu}$, where $g=\det g_{\mu \nu}$ is the determinant of the covariant metric. Because $\det f^{\mu \nu} = -c_{s}^{6}/V_{0}^{4}$ where $c_{s}:=\left( {V_{0}n_{0} / m} \right)^{1/2}$ is the local speed of sound in the WIBG, one finds that $\sqrt{-g}={n_{0}^{2}/ m^{2}c_{s}}$. Therefore, the contravariant and covariant expressions for the metric (written as 4$\times$4 matrices) are given by

\begin{equation}
g^{\mu\nu} = {m \over n_{0}c_{s}}\left[
\begin{array}{c|c}
  -1 & \left(v - {1 \over m}\nabla \theta_{0}\right) \\
  \hline
  \left(v - {1 \over m}\nabla \theta_{0}\right)^{T} & {V_{0}n_{0} \over m}\mathbb{I}_{3\times 3}-\left(v - {1 \over m}\nabla \theta_{0}\right)^{T}\left(v - {1 \over m}\nabla \theta_{0}\right)
\end{array}
\right] 
\label{eqn:contravar}
\end{equation}
and 
\begin{eqnarray}
g_{\mu\nu} 
&=&{n_{0}\over mc_{s}} \left[
\begin{array}{c|c}
  -c_{s}^{2}+\left(v - {1 \over m}\nabla \theta_{0}\right)\cdot\left(v - {1 \over m}\nabla \theta_{0}\right) & \left(v - {1 \over m}\nabla \theta_{0}\right) \\
  \hline
  \left(v - {1 \over m}\nabla \theta_{0}\right)^{T} & \mathbb{I}_{3\times 3}
\end{array}
\right]
\label{eqn:covar}
\end{eqnarray}
respectively.

All $n$-loop contributions to the action vanish for $n\ge 3$. The functional derivatives contributing to the two loop action scale as $\mathcal{O}(n_{0}(x)^{0})$ and are given by:
\begin{eqnarray}
{\delta^{3} S \over \delta n(x'',\tau '')\delta n(x',\tau') \delta\theta(x,\tau)}&=& 0 ,\nonumber \\
{\delta^{3} S \over \delta n(x'',\tau '')\delta \theta(x',\tau') \delta n(x,\tau)}&=& 0 ,\nonumber \\
{\delta^{3} S \over \delta n(x'',\tau '')\delta \theta(x',\tau') \delta n(x,\tau)}&=& 0 ,\nonumber \\
{\delta^{3} S \over \delta \theta (x'',\tau '')\delta n(x',\tau') \delta \theta(x,\tau)} &=& -{1\over m}\nabla \delta(x''-x)\cdot \nabla \delta (x-x') - {1\over m}\delta(x-x')\nabla^{2}\delta(x''-x) ,\nonumber \\ 
{\delta^{3} S \over \delta \theta(x'',\tau '')\delta \theta(x',\tau') \delta n(x,\tau)}&=&{1\over m}\nabla \delta(x''-x)\cdot \nabla \delta (x-x'), \nonumber \\ 
{\delta^{3} S \over \delta n(x'',\tau '')\delta \theta(x',\tau') \delta \theta(x,\tau)} &=& -{1\over m}\nabla \delta(x''-x)\cdot \nabla \delta (x-x') - {1\over m}\delta(x''-x)\nabla^{2}\delta(x'-x).
\label{eqn:usualthird}
\end{eqnarray}
\end{widetext}
There is a single nonvanishing two-loop contribution representing the scattering of two phase fluctuations to produce an amplitude fluctuation.  The contribution of this term to the effective action is $-{1\over 3!}\int_{[0,\beta \hbar]}\int_{\Omega}{1\over m}n_{d}\left( \nabla \theta_{d}\cdot \nabla \theta_{d} - \theta_{d}\nabla^{2}\theta_{d}\right)$. Integration over $n_{d}$ would produce an action that contains third-order and fourth-order derivatives of $\theta_{d}$. These terms should be included if one aims to deduce the short wavelength spectrum of the phase fluctuations, but can be neglected if one restricts to the long wavelength limit of the dynamics. Assuming that this restriction is made, the theory is exact at one-loop order.

By utilizing the imaginary time coherent state path integral, we can understand the effect of nonzero temperature on the effective ACS dynamics of $\theta_{d}$. From a physical perspective, it is clear that increasing the temperature of the nonrelativistic weakly imperfect Bose gas will destroy the analogue curved spacetime description of its phonon modes for two reasons: (1) at high temperatures, the spectrum of the gas approximates that of nonrelativistic free particles, so that there is no local Lorentz symmetry for any degree of freedom and (2) the fluctuations that propagate on the analogue curved spacetime are fluctuations of the phase of the bosonic field, which has a nonzero expectation value only at low temperatures. Mathematically, when the temperature is larger than any energy scale of the Hamiltonian, only the zeroth Matusbara frequency of the Fourier transformed phase fluctuation field $\tilde{\theta_{d}}(x,\omega_{n})$ contributes to the partition function \cite{tsvelikbook}. Therefore, there is no analogue curved spacetime in this case because the phase fluctuation field is time independent.

\bibliography{ACS12.bib}

\begin{thebibliography}{33}%
\makeatletter
\providecommand \@ifxundefined [1]{%
 \@ifx{#1\undefined}
}%
\providecommand \@ifnum [1]{%
 \ifnum #1\expandafter \@firstoftwo
 \else \expandafter \@secondoftwo
 \fi
}%
\providecommand \@ifx [1]{%
 \ifx #1\expandafter \@firstoftwo
 \else \expandafter \@secondoftwo
 \fi
}%
\providecommand \natexlab [1]{#1}%
\providecommand \enquote  [1]{``#1''}%
\providecommand \bibnamefont  [1]{#1}%
\providecommand \bibfnamefont [1]{#1}%
\providecommand \citenamefont [1]{#1}%
\providecommand \href@noop [0]{\@secondoftwo}%
\providecommand \href [0]{\begingroup \@sanitize@url \@href}%
\providecommand \@href[1]{\@@startlink{#1}\@@href}%
\providecommand \@@href[1]{\endgroup#1\@@endlink}%
\providecommand \@sanitize@url [0]{\catcode `\\12\catcode `\$12\catcode
  `\&12\catcode `\#12\catcode `\^12\catcode `\_12\catcode `\%12\relax}%
\providecommand \@@startlink[1]{}%
\providecommand \@@endlink[0]{}%
\providecommand \url  [0]{\begingroup\@sanitize@url \@url }%
\providecommand \@url [1]{\endgroup\@href {#1}{\urlprefix }}%
\providecommand \urlprefix  [0]{URL }%
\providecommand \Eprint [0]{\href }%
\providecommand \doibase [0]{http://dx.doi.org/}%
\providecommand \selectlanguage [0]{\@gobble}%
\providecommand \bibinfo  [0]{\@secondoftwo}%
\providecommand \bibfield  [0]{\@secondoftwo}%
\providecommand \translation [1]{[#1]}%
\providecommand \BibitemOpen [0]{}%
\providecommand \bibitemStop [0]{}%
\providecommand \bibitemNoStop [0]{.\EOS\space}%
\providecommand \EOS [0]{\spacefactor3000\relax}%
\providecommand \BibitemShut  [1]{\csname bibitem#1\endcsname}%
\let\auto@bib@innerbib\@empty
\bibitem [{\citenamefont {Barcel\'o}\ \emph {et~al.}(2011)\citenamefont
  {Barcel\'o}, \citenamefont {Liberati},\ and\ \citenamefont
  {Visser}}]{barcelorev}%
  \BibitemOpen
  \bibfield  {author} {\bibinfo {author} {\bibfnamefont {C.}~\bibnamefont
  {Barcel\'o}}, \bibinfo {author} {\bibfnamefont {S.}~\bibnamefont {Liberati}},
  \ and\ \bibinfo {author} {\bibfnamefont {M.}~\bibnamefont {Visser}},\
  }\bibfield  {title} {\enquote {\bibinfo {title} {{Analogue gravity}},}\
  }\href@noop {} {\bibfield  {journal} {\bibinfo  {journal} {Living Rev.
  Relativity}\ }\textbf {\bibinfo {volume} {14}},\ \bibinfo {pages} {3}
  (\bibinfo {year} {2011})}\BibitemShut {NoStop}%
\bibitem [{\citenamefont {Steinhauer}(2014)}]{steinhauer1}%
  \BibitemOpen
  \bibfield  {author} {\bibinfo {author} {\bibfnamefont {J.}~\bibnamefont
  {Steinhauer}},\ }\bibfield  {title} {\enquote {\bibinfo {title} {Observation
  of {self-amplifying} {Hawking} radiation in an analogue black-hole laser},}\
  }\href@noop {} {\bibfield  {journal} {\bibinfo  {journal} {Nat. Phys.}\
  }\textbf {\bibinfo {volume} {10}},\ \bibinfo {pages} {864} (\bibinfo {year}
  {2014})}\BibitemShut {NoStop}%
\bibitem [{\citenamefont {Corley}\ and\ \citenamefont
  {Jacobson}(1999)}]{BHlaser}%
  \BibitemOpen
  \bibfield  {author} {\bibinfo {author} {\bibfnamefont {S.}~\bibnamefont
  {Corley}}\ and\ \bibinfo {author} {\bibfnamefont {T.}~\bibnamefont
  {Jacobson}},\ }\bibfield  {title} {\enquote {\bibinfo {title} {Black hole
  lasers},}\ }\href {\doibase 10.1103/PhysRevD.59.124011} {\bibfield  {journal}
  {\bibinfo  {journal} {Phys. Rev. D}\ }\textbf {\bibinfo {volume} {59}},\
  \bibinfo {pages} {124011} (\bibinfo {year} {1999})}\BibitemShut {NoStop}%
\bibitem [{\citenamefont {Hung}\ \emph {et~al.}(2013)\citenamefont {Hung},
  \citenamefont {Gurarie},\ and\ \citenamefont {Chin}}]{Chin}%
  \BibitemOpen
  \bibfield  {author} {\bibinfo {author} {\bibfnamefont {C.-L.}\ \bibnamefont
  {Hung}}, \bibinfo {author} {\bibfnamefont {V.}~\bibnamefont {Gurarie}}, \
  and\ \bibinfo {author} {\bibfnamefont {C.}~\bibnamefont {Chin}},\ }\bibfield
  {title} {\enquote {\bibinfo {title} {{From cosmology to cold atoms:
  Observation of Sakharov oscillations in a quenched atomic superfluid}},}\
  }\href@noop {} {\bibfield  {journal} {\bibinfo  {journal} {Science}\ }\textbf
  {\bibinfo {volume} {341}},\ \bibinfo {pages} {1213} (\bibinfo {year}
  {2013})}\BibitemShut {NoStop}%
\bibitem [{\citenamefont {{Sakharov}}(1966)}]{Sakharov}%
  \BibitemOpen
  \bibfield  {author} {\bibinfo {author} {\bibfnamefont {A.~D.}\ \bibnamefont
  {{Sakharov}}},\ }\bibfield  {title} {\enquote {\bibinfo {title} {The initial
  stage of an expanding universe and the appearance of a nonuniform
  distribution of matter},}\ }\href@noop {} {\bibfield  {journal} {\bibinfo
  {journal} {J. Exp. Theor. Phys.}\ }\textbf {\bibinfo {volume} {22}},\
  \bibinfo {pages} {241} (\bibinfo {year} {1966})}\BibitemShut {NoStop}%
\bibitem [{\citenamefont {Jaskula}\ \emph {et~al.}(2012)\citenamefont
  {Jaskula}, \citenamefont {Partridge}, \citenamefont {Bonneau}, \citenamefont
  {Lopes}, \citenamefont {Ruaudel}, \citenamefont {Boiron},\ and\ \citenamefont
  {Westbrook}}]{Jaskula}%
  \BibitemOpen
  \bibfield  {author} {\bibinfo {author} {\bibfnamefont {J.-C.}\ \bibnamefont
  {Jaskula}}, \bibinfo {author} {\bibfnamefont {G.~B.}\ \bibnamefont
  {Partridge}}, \bibinfo {author} {\bibfnamefont {M.}~\bibnamefont {Bonneau}},
  \bibinfo {author} {\bibfnamefont {R.}~\bibnamefont {Lopes}}, \bibinfo
  {author} {\bibfnamefont {J.}~\bibnamefont {Ruaudel}}, \bibinfo {author}
  {\bibfnamefont {D.}~\bibnamefont {Boiron}}, \ and\ \bibinfo {author}
  {\bibfnamefont {C.~I.}\ \bibnamefont {Westbrook}},\ }\bibfield  {title}
  {\enquote {\bibinfo {title} {{Acoustic Analog to the Dynamical Casimir Effect
  in a Bose-Einstein Condensate}},}\ }\href@noop {} {\bibfield  {journal}
  {\bibinfo  {journal} {Phys. Rev. Lett.}\ }\textbf {\bibinfo {volume} {109}},\
  \bibinfo {pages} {220401} (\bibinfo {year} {2012})}\BibitemShut {NoStop}%
\bibitem [{\citenamefont {Fedichev}\ and\ \citenamefont {Fischer}(2004)}]{CPP}%
  \BibitemOpen
  \bibfield  {author} {\bibinfo {author} {\bibfnamefont {P.~O.}\ \bibnamefont
  {Fedichev}}\ and\ \bibinfo {author} {\bibfnamefont {U.~R.}\ \bibnamefont
  {Fischer}},\ }\bibfield  {title} {\enquote {\bibinfo {title}
  {``{C}osmological'' quasiparticle production in harmonically trapped
  superfluid gases},}\ }\href {\doibase 10.1103/PhysRevA.69.033602} {\bibfield
  {journal} {\bibinfo  {journal} {Phys. Rev. A}\ }\textbf {\bibinfo {volume}
  {69}},\ \bibinfo {pages} {033602} (\bibinfo {year} {2004})}\BibitemShut
  {NoStop}%
\bibitem [{\citenamefont {Korepin}\ \emph {et~al.}(1993)\citenamefont
  {Korepin}, \citenamefont {Izergin},\ and\ \citenamefont
  {Bogoliubov}}]{izergin}%
  \BibitemOpen
  \bibfield  {author} {\bibinfo {author} {\bibfnamefont {V.~E.}\ \bibnamefont
  {Korepin}}, \bibinfo {author} {\bibfnamefont {A.~G.}\ \bibnamefont
  {Izergin}}, \ and\ \bibinfo {author} {\bibfnamefont {N.~M.}\ \bibnamefont
  {Bogoliubov}},\ }\href@noop {} {\emph {\bibinfo {title} {{Quantum Inverse
  Scattering Method and Correlation Functions}}}}\ (\bibinfo  {publisher}
  {Cambridge University Press, Cambridge, England},\ \bibinfo {year}
  {1993})\BibitemShut {NoStop}%
\bibitem [{\citenamefont {Coleman}(1975)}]{colemanthirr}%
  \BibitemOpen
  \bibfield  {author} {\bibinfo {author} {\bibfnamefont {S.}~\bibnamefont
  {Coleman}},\ }\bibfield  {title} {\enquote {\bibinfo {title} {Quantum
  {sine-Gordon} equation as the massive {Thirring} model},}\ }\href@noop {}
  {\bibfield  {journal} {\bibinfo  {journal} {Phys. Rev. D}\ }\textbf {\bibinfo
  {volume} {11}},\ \bibinfo {pages} {2088} (\bibinfo {year}
  {1975})}\BibitemShut {NoStop}%
\bibitem [{\citenamefont {Barone}\ and\ \citenamefont
  {Patern\`{o}}(1982)}]{paterno}%
  \BibitemOpen
  \bibfield  {author} {\bibinfo {author} {\bibfnamefont {A.}~\bibnamefont
  {Barone}}\ and\ \bibinfo {author} {\bibfnamefont {G.}~\bibnamefont
  {Patern\`{o}}},\ }\href@noop {} {\emph {\bibinfo {title} {{Physics and
  Applications of the Josephson Effect}}}}\ (\bibinfo  {publisher} {Wiley, New
  York},\ \bibinfo {year} {1982})\BibitemShut {NoStop}%
\bibitem [{\citenamefont {Ustinov}\ \emph {et~al.}(1992)\citenamefont
  {Ustinov}, \citenamefont {Doderer}, \citenamefont {Huebener}, \citenamefont
  {Pedersen}, \citenamefont {Mayer},\ and\ \citenamefont {Oboznov}}]{ustinov}%
  \BibitemOpen
  \bibfield  {author} {\bibinfo {author} {\bibfnamefont {A.~V.}\ \bibnamefont
  {Ustinov}}, \bibinfo {author} {\bibfnamefont {T.}~\bibnamefont {Doderer}},
  \bibinfo {author} {\bibfnamefont {R.~P.}\ \bibnamefont {Huebener}}, \bibinfo
  {author} {\bibfnamefont {N.~F.}\ \bibnamefont {Pedersen}}, \bibinfo {author}
  {\bibfnamefont {B.}~\bibnamefont {Mayer}}, \ and\ \bibinfo {author}
  {\bibfnamefont {V.~A.}\ \bibnamefont {Oboznov}},\ }\bibfield  {title}
  {\enquote {\bibinfo {title} {{Dynamics of sine-Gordon Solitons in the Annular
  Josephson Junction}},}\ }\href@noop {} {\bibfield  {journal} {\bibinfo
  {journal} {Phys. Rev. Lett.}\ }\textbf {\bibinfo {volume} {69}},\ \bibinfo
  {pages} {1815} (\bibinfo {year} {1992})}\BibitemShut {NoStop}%
\bibitem [{\citenamefont {Gritsev}\ \emph {et~al.}(2007)\citenamefont
  {Gritsev}, \citenamefont {Polkovnikov},\ and\ \citenamefont
  {Demler}}]{demlersine}%
  \BibitemOpen
  \bibfield  {author} {\bibinfo {author} {\bibfnamefont {V.}~\bibnamefont
  {Gritsev}}, \bibinfo {author} {\bibfnamefont {A.}~\bibnamefont
  {Polkovnikov}}, \ and\ \bibinfo {author} {\bibfnamefont {E.}~\bibnamefont
  {Demler}},\ }\bibfield  {title} {\enquote {\bibinfo {title} {Linear response
  theory for a pair of coupled one-dimensional condensates of interacting
  atoms},}\ }\href@noop {} {\bibfield  {journal} {\bibinfo  {journal} {Phys.
  Rev. B}\ }\textbf {\bibinfo {volume} {75}},\ \bibinfo {pages} {174511}
  (\bibinfo {year} {2007})}\BibitemShut {NoStop}%
\bibitem [{\citenamefont {Neuenhahn}\ and\ \citenamefont
  {Marquardt}(2015)}]{marquardtcosmo}%
  \BibitemOpen
  \bibfield  {author} {\bibinfo {author} {\bibfnamefont {C.}~\bibnamefont
  {Neuenhahn}}\ and\ \bibinfo {author} {\bibfnamefont {F.}~\bibnamefont
  {Marquardt}},\ }\bibfield  {title} {\enquote {\bibinfo {title} {{Quantum
  simulation of expanding space-time with tunnel-coupled condensates}},}\
  }\href@noop {} {\bibfield  {journal} {\bibinfo  {journal} {New J. Phys.}\
  }\textbf {\bibinfo {volume} {17}},\ \bibinfo {pages} {125007} (\bibinfo
  {year} {2015})}\BibitemShut {NoStop}%
\bibitem [{\citenamefont {Negele}\ and\ \citenamefont {Orland}(1988)}]{negele}%
  \BibitemOpen
  \bibfield  {author} {\bibinfo {author} {\bibfnamefont {J.W.}\ \bibnamefont
  {Negele}}\ and\ \bibinfo {author} {\bibfnamefont {H.}~\bibnamefont
  {Orland}},\ }\href@noop {} {\emph {\bibinfo {title} {{Quantum Many-Particle
  Systems}}}}\ (\bibinfo  {publisher} {Addison-Wesley, Reading, MA},\ \bibinfo
  {year} {1988})\BibitemShut {NoStop}%
\bibitem [{\citenamefont {Griffin}\ \emph {et~al.}(2009)\citenamefont
  {Griffin}, \citenamefont {Nikuni},\ and\ \citenamefont
  {Zaremba}}]{zarembabook}%
  \BibitemOpen
  \bibfield  {author} {\bibinfo {author} {\bibfnamefont {A.}~\bibnamefont
  {Griffin}}, \bibinfo {author} {\bibfnamefont {T.}~\bibnamefont {Nikuni}}, \
  and\ \bibinfo {author} {\bibfnamefont {E.}~\bibnamefont {Zaremba}},\
  }\href@noop {} {\emph {\bibinfo {title} {{Bose-Einstein} Condensation at
  Finite Temperatures}}}\ (\bibinfo  {publisher} {Cambridge University Press,
  Cambridge, England},\ \bibinfo {year} {2009})\BibitemShut {NoStop}%
\bibitem [{Note1()}]{Note1}%
  \BibitemOpen
  \bibinfo {note} {This condition is one of four assumptions that, taken
  together, are sufficient for the derivation of a metric tensor that defines
  the local spacetime on which phase fluctuations of the WIBG propagate. The
  four assumptions are discussed in Appendix A.}\BibitemShut {Stop}%
\bibitem [{\citenamefont {Barcel\'o}\ \emph {et~al.}(2001)\citenamefont
  {Barcel\'o}, \citenamefont {Liberati},\ and\ \citenamefont
  {Visser}}]{BLVBEC}%
  \BibitemOpen
  \bibfield  {author} {\bibinfo {author} {\bibfnamefont {C.}~\bibnamefont
  {Barcel\'o}}, \bibinfo {author} {\bibfnamefont {S.}~\bibnamefont {Liberati}},
  \ and\ \bibinfo {author} {\bibfnamefont {M.}~\bibnamefont {Visser}},\
  }\bibfield  {title} {\enquote {\bibinfo {title} {{Analogue gravity from
  Bose-Einstein condensates}},}\ }\href
  {http://stacks.iop.org/0264-9381/18/i=6/a=312} {\bibfield  {journal}
  {\bibinfo  {journal} {Classical and Quantum Gravity}\ }\textbf {\bibinfo
  {volume} {18}},\ \bibinfo {pages} {1137} (\bibinfo {year}
  {2001})}\BibitemShut {NoStop}%
\bibitem [{\citenamefont {Evans}(2010)}]{evanspde}%
  \BibitemOpen
  \bibfield  {author} {\bibinfo {author} {\bibfnamefont {L.~C.}\ \bibnamefont
  {Evans}},\ }\href@noop {} {\emph {\bibinfo {title} {{Partial Differential
  Equations}}}}\ (\bibinfo  {publisher} {American Mathematical Society},\
  \bibinfo {year} {2010})\BibitemShut {NoStop}%
\bibitem [{\citenamefont {Bogoliubov}\ \emph {et~al.}(1995)\citenamefont
  {Bogoliubov}, \citenamefont {Kurbatov},\ and\ \citenamefont
  {Shumovsky}}]{bogo}%
  \BibitemOpen
  \bibfield  {author} {\bibinfo {author} {\bibfnamefont {N.~N.}\ \bibnamefont
  {Bogoliubov}}, \bibinfo {author} {\bibfnamefont {A.~M.}\ \bibnamefont
  {Kurbatov}}, \ and\ \bibinfo {author} {\bibfnamefont {A.~S.}\ \bibnamefont
  {Shumovsky}},\ }\href@noop {} {\emph {\bibinfo {title} {Quantum and Classical
  Statistical Mechanics}}},\ Classics of Soviet Mathematics\ (\bibinfo
  {publisher} {Gordon and Breach, London},\ \bibinfo {year} {1995})\BibitemShut
  {NoStop}%
\bibitem [{\citenamefont {Zagrebnov}\ and\ \citenamefont
  {Bru}(2001)}]{zagrebnov}%
  \BibitemOpen
  \bibfield  {author} {\bibinfo {author} {\bibfnamefont {V.A.}\ \bibnamefont
  {Zagrebnov}}\ and\ \bibinfo {author} {\bibfnamefont {J.-B.}\ \bibnamefont
  {Bru}},\ }\bibfield  {title} {\enquote {\bibinfo {title} {The {Bogoliubov}
  model of weakly imperfect {Bose} gas},}\ }\href@noop {} {\bibfield  {journal}
  {\bibinfo  {journal} {Physics Rep.}\ }\textbf {\bibinfo {volume} {350}},\
  \bibinfo {pages} {291} (\bibinfo {year} {2001})}\BibitemShut {NoStop}%
\bibitem [{\citenamefont {Lieb}\ and\ \citenamefont
  {Seiringer}(2005)}]{liebcnumber}%
  \BibitemOpen
  \bibfield  {author} {\bibinfo {author} {\bibfnamefont {E.~H.}\ \bibnamefont
  {Lieb}}\ and\ \bibinfo {author} {\bibfnamefont {R.}~\bibnamefont
  {Seiringer}},\ }\bibfield  {title} {\enquote {\bibinfo {title}
  {{Justification of c-Number Substitutions in Bosonic {Hamiltonians}}},}\
  }\href@noop {} {\bibfield  {journal} {\bibinfo  {journal} {Phys. Rev. Lett.}\
  }\textbf {\bibinfo {volume} {94}},\ \bibinfo {pages} {080401} (\bibinfo
  {year} {2005})}\BibitemShut {NoStop}%
\bibitem [{\citenamefont {Abrikosov}\ \emph {et~al.}(1965)\citenamefont
  {Abrikosov}, \citenamefont {Gor'kov},\ and\ \citenamefont
  {Dzyaloshinski\v\i}}]{abrikosov}%
  \BibitemOpen
  \bibfield  {author} {\bibinfo {author} {\bibfnamefont {A.~A.}\ \bibnamefont
  {Abrikosov}}, \bibinfo {author} {\bibfnamefont {L.~P.}\ \bibnamefont
  {Gor'kov}}, \ and\ \bibinfo {author} {\bibfnamefont {I.~E.}\ \bibnamefont
  {Dzyaloshinski\v\i}},\ }\href@noop {} {\emph {\bibinfo {title} {{Methods of
  Quantum Field Theory in Statistical Physics}}}}\ (\bibinfo  {publisher}
  {Pergamon Press, New York},\ \bibinfo {year} {1965})\BibitemShut {NoStop}%
\bibitem [{\citenamefont {Shi}\ and\ \citenamefont
  {Griffin}(1998)}]{griffinshi}%
  \BibitemOpen
  \bibfield  {author} {\bibinfo {author} {\bibfnamefont {H.}~\bibnamefont
  {Shi}}\ and\ \bibinfo {author} {\bibfnamefont {A.}~\bibnamefont {Griffin}},\
  }\bibfield  {title} {\enquote {\bibinfo {title} {Finite-temperature
  excitations in a dilute {Bose}-condensed gas},}\ }\href@noop {} {\bibfield
  {journal} {\bibinfo  {journal} {Phys. Rep.}\ }\textbf {\bibinfo {volume}
  {304}},\ \bibinfo {pages} {1} (\bibinfo {year} {1998})}\BibitemShut {NoStop}%
\bibitem [{\citenamefont {Volkoff}\ and\ \citenamefont
  {Whaley}(2014)}]{volkoff}%
  \BibitemOpen
  \bibfield  {author} {\bibinfo {author} {\bibfnamefont {T.~J.}\ \bibnamefont
  {Volkoff}}\ and\ \bibinfo {author} {\bibfnamefont {K.~B.}\ \bibnamefont
  {Whaley}},\ }\bibfield  {title} {\enquote {\bibinfo {title} {Measurement- and
  comparison-based sizes of {Schr\"{o}dinger} cat states of light},}\
  }\href@noop {} {\bibfield  {journal} {\bibinfo  {journal} {Phys. Rev. A}\
  }\textbf {\bibinfo {volume} {91}},\ \bibinfo {pages} {012122} (\bibinfo
  {year} {2014})}\BibitemShut {NoStop}%
\bibitem [{\citenamefont {Dodonov}\ \emph {et~al.}(1974)\citenamefont
  {Dodonov}, \citenamefont {Malkin},\ and\ \citenamefont {Man'ko}}]{dodonov}%
  \BibitemOpen
  \bibfield  {author} {\bibinfo {author} {\bibfnamefont {V.~V.}\ \bibnamefont
  {Dodonov}}, \bibinfo {author} {\bibfnamefont {I.~A.}\ \bibnamefont {Malkin}},
  \ and\ \bibinfo {author} {\bibfnamefont {V.~I.}\ \bibnamefont {Man'ko}},\
  }\bibfield  {title} {\enquote {\bibinfo {title} {Even and odd coherent states
  and excitations of a singular oscillator},}\ }\href@noop {} {\bibfield
  {journal} {\bibinfo  {journal} {Physica}\ }\textbf {\bibinfo {volume} {72}},\
  \bibinfo {pages} {597} (\bibinfo {year} {1974})}\BibitemShut {NoStop}%
\bibitem [{\citenamefont {Cochrane}\ \emph {et~al.}(1999)\citenamefont
  {Cochrane}, \citenamefont {Milburn},\ and\ \citenamefont
  {Munro}}]{milburncatcode}%
  \BibitemOpen
  \bibfield  {author} {\bibinfo {author} {\bibfnamefont {P.~T.}\ \bibnamefont
  {Cochrane}}, \bibinfo {author} {\bibfnamefont {G.~J.}\ \bibnamefont
  {Milburn}}, \ and\ \bibinfo {author} {\bibfnamefont {W.~J.}\ \bibnamefont
  {Munro}},\ }\bibfield  {title} {\enquote {\bibinfo {title} {Macroscopically
  distinct quantum-superposition states as a bosonic code for amplitude
  damping},}\ }\href@noop {} {\bibfield  {journal} {\bibinfo  {journal} {Phys.
  Rev. A}\ }\textbf {\bibinfo {volume} {59}},\ \bibinfo {pages} {2631}
  (\bibinfo {year} {1999})}\BibitemShut {NoStop}%
\bibitem [{\citenamefont {Raghavan}\ \emph {et~al.}(1999)\citenamefont
  {Raghavan}, \citenamefont {Smerzi}, \citenamefont {Fantoni},\ and\
  \citenamefont {Shenoy}}]{Raghavan}%
  \BibitemOpen
  \bibfield  {author} {\bibinfo {author} {\bibfnamefont {S.}~\bibnamefont
  {Raghavan}}, \bibinfo {author} {\bibfnamefont {A.}~\bibnamefont {Smerzi}},
  \bibinfo {author} {\bibfnamefont {S.}~\bibnamefont {Fantoni}}, \ and\
  \bibinfo {author} {\bibfnamefont {S.~R.}\ \bibnamefont {Shenoy}},\ }\bibfield
   {title} {\enquote {\bibinfo {title} {{Coherent oscillations between two
  weakly coupled Bose-Einstein condensates: Josephson effects,
  $\ensuremath{\pi}$ oscillations, and macroscopic quantum self-trapping}},}\
  }\href {\doibase 10.1103/PhysRevA.59.620} {\bibfield  {journal} {\bibinfo
  {journal} {Phys. Rev. A}\ }\textbf {\bibinfo {volume} {59}},\ \bibinfo
  {pages} {620} (\bibinfo {year} {1999})}\BibitemShut {NoStop}%
\bibitem [{\citenamefont {Volkoff}(2015)}]{volkoffcat}%
  \BibitemOpen
  \bibfield  {author} {\bibinfo {author} {\bibfnamefont {T.~J.}\ \bibnamefont
  {Volkoff}},\ }\bibfield  {title} {\enquote {\bibinfo {title} {Nonclassical
  properties and quantum resources of hierarchical photonic superposition
  states},}\ }\href@noop {} {\bibfield  {journal} {\bibinfo  {journal} {J. Exp.
  Theor. Phys.}\ }\textbf {\bibinfo {volume} {121}},\ \bibinfo {pages} {770}
  (\bibinfo {year} {2015})}\BibitemShut {NoStop}%
\bibitem [{\citenamefont {Callan}\ and\ \citenamefont
  {Coleman}(1977)}]{coleman}%
  \BibitemOpen
  \bibfield  {author} {\bibinfo {author} {\bibfnamefont {C.~G.}\ \bibnamefont
  {Callan}}\ and\ \bibinfo {author} {\bibfnamefont {S.}~\bibnamefont
  {Coleman}},\ }\bibfield  {title} {\enquote {\bibinfo {title} {Fate of the
  false vacuum. {II.} {First} quantum corrections},}\ }\href {\doibase
  10.1103/PhysRevD.16.1762} {\bibfield  {journal} {\bibinfo  {journal} {Phys.
  Rev. D}\ }\textbf {\bibinfo {volume} {16}},\ \bibinfo {pages} {1762}
  (\bibinfo {year} {1977})}\BibitemShut {NoStop}%
\bibitem [{\citenamefont {Ginzburg}\ and\ \citenamefont
  {Pitaevski\v\i}(1958)}]{pitaevskii}%
  \BibitemOpen
  \bibfield  {author} {\bibinfo {author} {\bibfnamefont {V.~L.}\ \bibnamefont
  {Ginzburg}}\ and\ \bibinfo {author} {\bibfnamefont {L.~P.}\ \bibnamefont
  {Pitaevski\v\i}},\ }\bibfield  {title} {\enquote {\bibinfo {title} {On the
  theory of superfluidity},}\ }\href@noop {} {\bibfield  {journal} {\bibinfo
  {journal} {J. Exp. Theor. Phys.}\ }\textbf {\bibinfo {volume} {7}},\ \bibinfo
  {pages} {858} (\bibinfo {year} {1958})}\BibitemShut {NoStop}%
\bibitem [{\citenamefont {Fischer}\ and\ \citenamefont
  {Visser}(2003)}]{fischervisser1}%
  \BibitemOpen
  \bibfield  {author} {\bibinfo {author} {\bibfnamefont {U.~R.}\ \bibnamefont
  {Fischer}}\ and\ \bibinfo {author} {\bibfnamefont {M.}~\bibnamefont
  {Visser}},\ }\bibfield  {title} {\enquote {\bibinfo {title} {On the spacetime
  curvature experienced by quasiparticle excitations in the
  {Painlev\'e-Gullstrand} effective geometry},}\ }\href@noop {} {\bibfield
  {journal} {\bibinfo  {journal} {Ann. Phys. (N.Y.)}\ }\textbf {\bibinfo
  {volume} {304}},\ \bibinfo {pages} {22} (\bibinfo {year} {2003})}\BibitemShut
  {NoStop}%
\bibitem [{\citenamefont {Fischer}\ and\ \citenamefont
  {Visser}(2002)}]{fischervisser2}%
  \BibitemOpen
  \bibfield  {author} {\bibinfo {author} {\bibfnamefont {U.~R.}\ \bibnamefont
  {Fischer}}\ and\ \bibinfo {author} {\bibfnamefont {M.}~\bibnamefont
  {Visser}},\ }\bibfield  {title} {\enquote {\bibinfo {title} {{Riemannian
  Geometry of Irrotational Vortex Acoustics}},}\ }\href@noop {} {\bibfield
  {journal} {\bibinfo  {journal} {Phys. Rev. Lett.}\ }\textbf {\bibinfo
  {volume} {88}},\ \bibinfo {pages} {110201} (\bibinfo {year}
  {2002})}\BibitemShut {NoStop}%
\bibitem [{\citenamefont {Tsvelik}(2003)}]{tsvelikbook}%
  \BibitemOpen
  \bibfield  {author} {\bibinfo {author} {\bibfnamefont {A.M.}\ \bibnamefont
  {Tsvelik}},\ }\href@noop {} {\emph {\bibinfo {title} {{Quantum Field Theory
  in Condensed Matter Physics}}}}\ (\bibinfo  {publisher} {Cambridge University
  Press, Cambridge, England},\ \bibinfo {year} {2003})\BibitemShut {NoStop}%
\end{thebibliography}%

\end{document}